\documentclass[12pt]{iopart}
\usepackage{graphicx}

\usepackage{array}
\usepackage[numbers,sort&compress]{natbib}
\usepackage{pifont}
\usepackage{xcolor}

\usepackage{makecell}
\usepackage[justification=raggedright]{caption}
\usepackage{comment}
\usepackage{amssymb}
\begin{document}
\newcolumntype{C}[1]{>{\centering\arraybackslash}p{#1}}
\title[]{Pinpointing energy transfer mechanisms in the quenching process of microwave air plasma}

\author{Q. Shen$^1$, A. Pikalev$^{1,2}$, F.J.J. Peeters$^3$, V. Guerra$^2$, M.C.M. van de Sanden$^{1,4,*}$ }

\address{1. Dutch Institute for Fundamental Energy Research, Eindhoven, The Netherlands}
\address{2. Instituto de Plasmas e Fusão Nuclear, Instituto Superior Técnico, Universidade de Lisboa, Lisboa, Portugal}

\address{3. Leyden Jar Company, Eindhoven, The Netherlands}
\address{4. Department of Applied Physics, Eindhoven Institute of Renewable Energy Systems, Eindhoven University of Technology, Eindhoven, The Netherlands}

\ead{M.C.M.vandeSanden@differ.nl}
\vspace{10pt}
\newcommand\myworries[1]{\textcolor{black}{#1}}
\newcommand\secondround[1]{\textcolor{black}{#1}}
\newcommand\thirdround[1]{\textcolor{black}{#1}}

\newcommand\fourround[1]{\textcolor{black}{#1}}
\newcommand\fiveround[1]{\textcolor{black}{#1}}

\begin{abstract}
  A time-dependent multi-temperature quenching model \secondround{at} atmospheric pressure, incorporating chemical and vibrational kinetics, is introduced. The model provides insights into the pathways of NO$_x$ formation and destruction in the downstream region of a microwave air plasma. \fourround{The} relaxation of the temperatures during the forced cooling trajectory \fourround{by the wall} is modelled. A Continuous Stirred Tank Reactor \secondround{model} and \secondround{a} Plug Flow Reactor model represent the plasma and quenching regions, respectively. \myworries{For the non-thermal \secondround{conditions,} where gas and vibrational temperatures differ, \secondround{most reaction rate coefficients, except those obtained from molecular dynamics methods, are determined based on a} generalized Fridman-Macheret scheme}. \secondround{The energy transfer channels involved in the quenching process are tracked across different time scales.} By varying the gas temperature in the plasma region and the cooling rate, the reaction pathways for the NO$_x$ synthesis mechanism are analysed. This research \fourround{provides a first step for} the further advancement and optimisation of plasma reactors for efficient NO$_x$ production.
\end{abstract}

\section{Introduction}

\secondround{Limited} availability of natural nitrogen-based fertilizer has restrained agricultural productivity until the advent of the Haber-Bosch (H-B) process. Nowadays, the H-B process fixes approximately 130 million tons of nitrogen annually, sustaining over 40\% of the global population \cite{patil2015plasma, kelly2021nitrogen}. However, this process is currently one of the largest global energy consumers and greenhouse gas emitters, accounting for about 1.2\% of the global anthropogenic CO$_2$ emissions, prompting researchers to recommend alternative Nitrogen Fixation (NF) methods \cite{smith2020current}. \secondround{Although major strides have been made to electrify the H-B process, it continues to rely heavily on fossil fuels and remains inflexible }\cite{rouwenhorst2021birkeland}. \secondround{Recently, plasma technology has attracted more and more attention, as it can be easily switched on and off,  making it compatible with fluctuating renewable energy sources such as solar or wind power \cite{wang2017nitrogen, winter2021n2}.} Consequently, the plasma NF process has the potential to significantly reduce CO$_2$ emissions \secondround{compared to} traditional methods, paving the way for sustainable NF with net-zero emissions in the future \cite{abdelaziz2023toward, li2022direct, jardali2021no}.

Several articles have discussed the underlying NO$_x$ synthesis \secondround{mechanisms in}  different plasma types \cite{abdelaziz2023toward, wang2017nitrogen, fridman2008plasma}. The most \secondround{energy-efficient} pathway for NO synthesis involves \secondround{stimulating non-thermal processes} through vibrational excitation, which is reported to have the potential to achieve a theoretical power consumption limit of 0.1~MJ/mol N$^{-1}$, \textit{i.e.,} 2.5 times lower than \secondround{that of} the H-B process \cite{liu2024plasma}. The highest NO production (14\%) with the lowest energy cost (0.28~MJ/mol N$^{-1}$) was achieved by Asisov \textit{et al.} employing a low-pressure (10$-$100 Torr) electron cyclotron resonance microwave (MW) discharge with a magnetic field and cryogenically cooled reactor wall \cite{asisov1980high, fridman2008plasma}. However, neither the energy consumed to sustain the low pressure in the reactor, nor the energy required for \secondround{cooling the wall is} included in the energy cost calculation \cite{muzammil2021novel, rouwenhorst2021birkeland}. 
For other typical non-thermal plasmas, \secondround{like} dielectric barrier discharges, the energetic electrons, excited by the \secondround{high} reduced electric field (above 100 Td), \thirdround{spend} too much energy on electronic excitation, ionisation, and dissociation, instead of vibrational excitation, resulting in a higher energy cost (18~MJ/mol N$^{-1}$) and low NO$_x$ production \cite{bogaerts2018plasma}.


Warm plasma is another promising option for NO$_x$ synthesis, because it can provide \secondround{a} high ionisation \secondround{degree}, coupled with non-thermal effects (\textit{i.e.,} selectively populate certain degrees of freedom, such as vibrationally excited states) \cite{snoeckx2017plasma}. \secondround{Although} the translational temperature in warm plasma is much lower than the electron temperature ($\sim$ 1$-$2 eV), it is \secondround{still} significantly higher than room temperature, \secondround{reaching} several thousand Kelvin \cite{bogaerts2018plasma}. Kelly \textit{et al.} \cite{kelly2021nitrogen} presented an atmospheric pressure MW plasma stabilised by a vortex gas flow, achieving an energy cost of 2~MJ/mol N$^{-1}$ and \secondround{a} NO$_x$ production of 3.8\%. Vervloessem \textit{et al.} \cite{vervloessem2020plasma} reported NO$_x$ formation of 1.5\% at an energy cost of 3.6~MJ/mol N$^{-1}$ based on \secondround{a} gliding arc (GA) plasmatron at the same pressure. A lower energy cost (1.8~MJ/mol N$^{-1}$) was achieved by Tsonev \textit{et~al.}  \cite{tsonev2023nitrogen} using a rotating GA plasma reactor at elevated pressure. \secondround{Still}, the energy cost of NO$_x$ synthesis in warm plasma \secondround{remains high}, because a high \secondround{gas} temperature is required \secondround{to} break the strong N-N chemical bonds, while NO \secondround{is unstable} at these elevated temperatures \cite{wang2017nitrogen}. \secondround{Furthermore}, effective quenching is necessary to prevent the NO$_x$ generated in the plasma zone from decomposing back into N$_2$ and O$_2$ \cite{fridman2008plasma}. \secondround{There is ample literature debating NO$_x$ formation and CO$_2$ dissociation mechanisms in plasmas, for sustainable chemistry applications. As it is also largely debated, the bottlenecks of those processes are the product quenching and separation of products \cite{van2022effusion, majeed2024effect,van2024effluent,mercer2023post}.} Van Alphen \textit{et al.} \cite{van2022effusion} indicated an 8\% enhancement in NO$_x$ concentration using \secondround{a} so-called \(\textquoteleft\)effusion nozzle' in \secondround{a} rotating GA plasma reactor. Similarly, Majeed \textit{et al.} \cite{majeed2024effect} found that NO$_x$ production increased by 12.4\% when \secondround{the mixture was mixed with a cold gas} after the plasma zone in the same reactor. Therefore, exploring the underlying mechanisms in the quenching region is crucial for \secondround{both future plasma reactor development and energy cost optimization}. 


Due to the short lifetimes of many intermediate products generated by plasma, \secondround{a} full understanding of NO$_x$ synthesis through experiments is extremely demanding \cite{liu2024plasma}. An accurate and comprehensive theoretical model allows researchers to gain a deeper and more quantitative knowledge of the \secondround{plasma} NF process \cite{esposito2022relevance}. Altin \textit{et al}. \cite{altin2022energy} developed a zero-dimensional (0D) chemistry model to \secondround{investigate} vibrational excitation in the core of pure N$_2$ \secondround{MW} plasmas, \secondround{with} pressure varying from 50 to 400 mbar. In a follow-up paper, a one-dimensional (1D) radial fluid model revealed the gas heating mechanism of the plasma core and its outer region in pure N$_2$ \secondround{MW} plasma \cite{altin2024spatio}. Based on a 0D chemical kinetics model, Wang \textit{et al}. \cite{wang2017nitrogen} illustrated the NO$_x$ synthesis in a GA plasma at atmospheric pressure. The results revealed that the vibrational excitation of N$_2$ can lower the high energy barrier for N$_2$ dissociation, thereby reducing the energy cost of the NF process. In another 0D chemical kinetics model for \secondround{a} reverse-vortex flow GA plasmatron at atmospheric pressure, Vervloessem \textit{et al}. \cite{vervloessem2020plasma} reported that the energy cost \secondround{could potentially be reduced} to 0.5 MJ/mol N$^{-1}$ by preventing the transfer of vibrational energy from N$_2$ to O$_2$. \myworries{The strong non-equilibrium effects in high-enthalpy gas \secondround{flows} expanding through nozzles have been investigated. Kustova \textit{et~al.} \cite{kustova2002non} examined how different vibrational distributions affect heat transfer and diffusion in expanding nozzle flows using various kinetic models. \secondround{Shizgal \textit{et al.} \cite{shizgal1996vibrational} found that the forward rates of O$_2$–O$_2$ and O$_2$–O dissociation reactions in hypersonic nozzle expansion significantly deviate from equilibrium due to vibrational nonequilibrium effects. The calculations showed that near the throat, all vibrational levels contribute to the forward reaction rates, whereas farther downstream, the highest vibrational levels dominate.} Cutler \textit{et al.} \cite{cutler2012measurement} experimentally and computationally investigated vibrational and rotational temperatures in a Mach 2 nozzle, showing vibrational freezing effects and the role of steam in promoting thermal equilibrium.}

In this work, we concentrate on the quenching of \secondround{air activated} by warm plasma using a time-dependent multi-temperature quenching model, which is coupled with chemical and vibrational kinetics under atmospheric pressure, \secondround{allowing for} the study of the underlying mechanisms. Overall, the model provides \fourround{a first step for} understanding the quenching process in the downstream afterglow of a MW air plasma, revealing the intricate balance \secondround{of} energy input, cooling rates, and chemical-kinetical reactions. This insight is crucial for optimizing NO$_x$ production while minimizing energy costs, laying the groundwork for further advancements in plasma-based nitrogen fixation technologies.

\section{Model description}

\begin{figure}[h]
\centering
\includegraphics[width=1\linewidth]{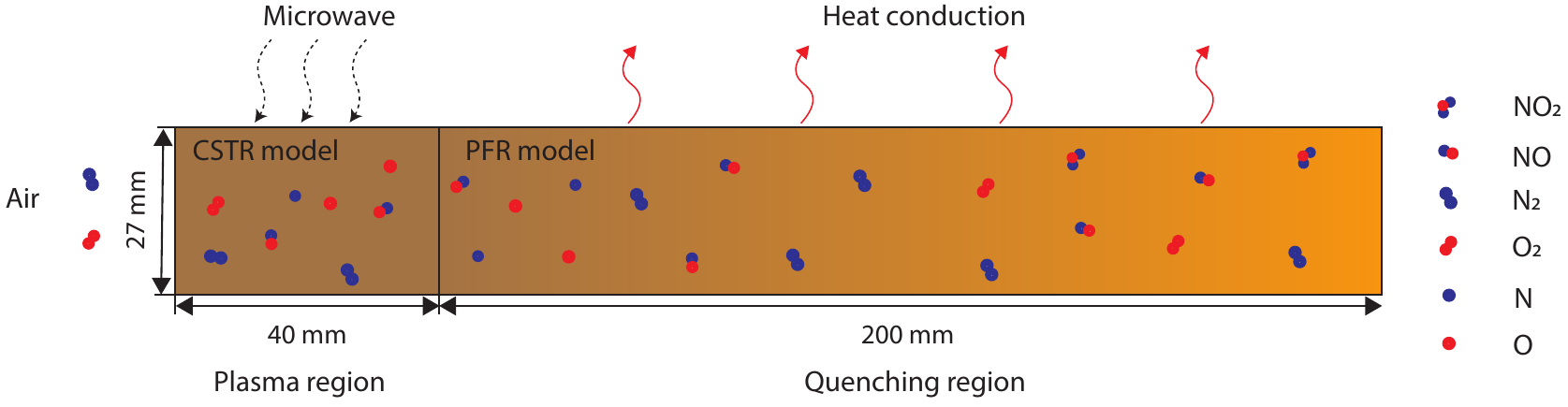}
\caption{ Sketch of the complete model in this work.  }
\label{skecth picture}
\end{figure}

In order to better understand the NO$_x$ formation and energy exchange mechanisms in the MW plasma, a 0D Continuous Stirred Tank Reactor (CSTR) model coupled with a 1D axial Plug Flow Reactor (PFR) model \secondround{is employed} to represent the plasma region and quenching region, respectively, as shown in Figure \ref{skecth picture}. During the discharge, 10 slm air (79\% N$_2$ and 21\% O$_2$) \secondround{undergoes} numerous physical and chemical processes, including electronic and vibrational excitation, ionisation, \secondround{dissociation}, and recombination \cite{liu2024plasma}. Since experiments using Raman spectroscopy have shown that the vibrational temperature is very close to the rotational temperature in the core of microwave air plasma at \thirdround{650} mbar \cite{tatar2024analysis}, it is valid to assume that the \secondround{gas mixture} is in thermal equilibrium in the plasma region at atmospheric pressure, and \secondround{can} be represented by a CSTR model. \secondround{The residence time in the CSTR model is a critical factor affecting the performance \cite{bongers2017plasma, kozak2014splitting}, particularly at low pressure. The length of the CSTR— the parameter governing the residence time— is fixed at 4~cm. Because the gas mixture reaches chemical equilibrium very fast under atmospheric \fourround{pressure}, a sensitivity test has confirmed that the length of the CSTR has a limited effect on the result when the temperature in the CSTR ($T_{CSTR}$) is beyond 2500~K (Figure \ref{fig:S1} in Supplementary material). Therefore, it is reasonable to assume that the initial gas reaches the chemical equilibrium \fourround{based on the gas temperature} in the CSTR model. Note that the calculations using the CSTR, which simulate the microwave plasma, only provide the initial conditions for our analysis of the quenching process in the afterglow and do not influence the analysis of energy transfer during quenching. }

The PFR model is represented by a series of small CSTRs connected in sequence. This approach assumes that the quenching region can be approximated by dividing it into \secondround{numerous} smaller compartments \cite{kee2005chemically}. Each CSTR represents a small volume within the PFR, where perfect mixing is assumed within that compartment, but not between different compartments. \secondround{The gas mixture entering the first CSTR has the specified inlet composition from the plasma region, while for all subsequent CSTRs, the inlet composition is fixed to match the reactor immediately upstream} \cite{cantera}. The PFR model typically assumes the fluid is perfectly mixed in the radial direction, which means that the \secondround{density profiles} of each component are uniform across the direction perpendicular to the axis \cite{kotov2019plug}. In a PFR model, where diffusion is absent, upstream reactors are unaffected by downstream reactors. Consequently, the problem can be solved by sequentially progressing from the first reactor to the last, integrating each into a steady state \cite{cantera}. The residence time \secondround{of} each step depends on the mass flow rate and \secondround{the} 0D reactor volume. By increasing the number of CSTRs, the model can capture the gradients in concentration or temperature across the reactor length. In this work, the length of each compartment in the PFR is variable. When any temperature difference—whether in gas temperature or in the vibrational temperatures of N$_2$ and O$_2$—between the current and previous compartment falls below 1 K or exceeds 10 K, the model recalculates the current step by adjusting the reactor length until the result meets the specified criteria, thus improving computational efficiency. The simulation continues until the cumulative length of all compartments reaches or exceeds 20 cm, \secondround{after} which all active species are depleted under \secondround{all investigated} conditions (the reaction 2NO+O$_2$\(\rightarrow\)2NO$_2$ is active at longer times and is discussed separately). Consequently, the number of compartments in the PFR reactor varies under different conditions. Moreover, because species \secondround{concentrations at atmospheric pressure} in the MW reactor \secondround{are} primarily determined by gas-phase chemistry \cite{vialetto2022charged}, surface reactions are not included in the model.

Since \secondround{the} translational relaxation time is shorter than the \secondround{vibrational-translational} relaxation time, apart from the plasma region, fast cooling in the quenching region is another possible way to achieve non-thermal effects, which might benefit NO production. The vibrational modes of N$_2$ and O$_2$ are assumed to follow \secondround{a} Boltzmann distribution controlled by their vibrational temperatures. As a result, nearly all reactions that \secondround{involve} O$_2$ or N$_2$ as \secondround{a} reactant can be vibrationally stimulated (\(i.e.,\) reactions X$_1^f$\(-\)X$_5^f$ in Table~\ref{tab:reactions}). The model is implemented \secondround{using} Cantera, an open-source Python package that solves chemical kinetics, thermodynamics, and transport processes \cite{cantera}.  

\thirdround{Several aspects demonstrate the accuracy and reliability of the present model. Firstly, the accuracy of the generalized Fridman-Macheret method used in this work has been proved by comparison with Quasi-Classical Trajectory (QCT) results \cite{esposito2006n,esposito2008o2}. Secondly, the predicted molar fractions with sufficient residence time in the CSTR show excellent agreement with those obtained from chemical equilibrium \cite{d2008thermodynamic}. Thirdly, under low cooling rate conditions, the model correctly predicts the absence of non-thermal equilibrium effects, consistent with experimental observations even at lower pressures \cite{tatar2024analysis}. Lastly, in high-enthalpy nozzle flows, a higher vibrational temperature is observed compared to the gas temperature, aligning with the prediction of the present model that stronger non-thermal effects persist under high cooling rates
\cite{kustova2002non, shizgal1996vibrational, cutler2012measurement}.}

\subsection{Chemistry set}

\secondround{Given that minimal} N$_2$O, N$_2$O$_3$, \secondround{or} N$_2$O$_4$ \secondround{is} observed in the experiments \cite{kelly2021nitrogen}, these species \secondround{are omitted from} the reaction model. Table \ref{tab:reactions} lists all chemical reactions \secondround{included in the model}. For the reactions X$_2^f$ and X$_5^f$, the thermal reaction rate coefficients are calculated by the molecular dynamics (MD) data \cite{armenise2021n+,armenise2023low, esposito2017reactive}:
\begin{equation}
\label{eq:MD}
    k= \sum_v f_v(T_g) \sum_w k_{v,w}(T_g)
\end{equation}
where \(f_v(T_g)\) is the Boltzmann factor of O$_2$ or N$_2$ \cite{shen2024two}, \(k_{v,w}(T_g)\) is the state-specific rate coefficient for \(v^{th}\) vibrational state of N$_2$ or O$_2$ as the reactant, and  \(w^{th}\) level of NO as the product. More information can be found in the work of Armenise \textit{et al.} \cite{armenise2021n+,armenise2023low, esposito2017reactive}.
\myworries{The thermal rate coefficients of the reaction O$_2$+O$_2$/O in reaction \(X_1^f\) and N$_2$+N$_2$/N in reaction \(X_4^f\) are calculated by the QCT data \cite{esposito2008o2, andrienko2017state,esposito2006n}:}
\begin{equation}
\label{eq:QCT}
   \myworries{ k= \sum_v f_v(T_g) k_{v}(T_g)}
\end{equation}

For the other reactions in Table \ref{tab:reactions}, all the forward rate coefficients (X$^f$) in thermal equilibrium are calculated using the modified Arrhenius form \cite{cantera}:

\begin{equation}
{k=AT_g^b \exp \left(-\frac{E_a}{k_BT_g}   \right)} 
\end{equation}
where \(k\) [m$^3$s$^{-1}$] is the reaction rate coefficient, \(A\) [m$^3$s$^{-1}$] is the pre-exponential factor, $k_B$ is \thirdround{the} Boltzmann constant, \(b\) is the temperature exponent, \(T_g\) [K] is the gas temperature, and \(E_a\) [eV] is the activation energy.

\myworries{All rate coefficients for the backward reactions are determined by detailed balance, (\textit{i.e.,} the net rate-of-progress of all reactions is zero \cite{vialetto2022charged}). The relation between the forward and backward rate coefficients is expressed by:}
\begin{equation}
{\myworries{K_{eq,i}\equiv \frac{k_{f,i}}{k_{b,i}}} }
\end{equation}
\myworries{where \(K_{eq,i}\) is the \secondround{equilibrium constant of the reaction \textit{i}}, calculated using the formula known from thermodynamics \cite{kotov2023validation}:}

\begin{equation}
{\secondround{K_{eq,i}= \left(\frac{P_{ref}}{k_BT_g}\right) ^{\sum_j a_{ij}}  \exp \left(-\frac{\sum_j a_{ij}  G_{ij}^{ref}(T_g)}{RT_g}\right)    } } 
\end{equation}
\secondround{where \( a_{ij}\) is the stoichiometric factors of species \textit{j} (positive for products and negative for reactants) of each chemical component in the corresponding process, the sums are taken over all components that take part in the given reaction; \(G_{ij}^{ref}\) [J\(\cdot\)mol$^{-1}$] are the molar Gibbs energies of components at the reference pressure \(P_{ref}\) [Pa]; \textit{R} is gas constant.}

\begin{table}[ht]
\caption{List of thermal chemistry reactions. All the rate coefficients for the backward reactions are determined by the principle of detailed balance.}
\centering
\label{tab:reactions}
\begin{tabular}{cccccccccccccc}
 \hline
 No. & Reaction & \(A\, (\mathrm{m}^3\mathrm{s}^{-1})\) & \(b\) & \(E_a \, (\mathrm{eV})\) & \(\alpha\)  & Ref \\
 \hline
 \textcolor{blue}{$^a$}X$_1^f$ & \(\mathrm{O}_2 + \mathrm{M} \rightarrow \mathrm{O} + \mathrm{O} + \mathrm{M}\) & \(3.32 \times 10^{-9}\) & -1.5 & 5.11 & 1 & \cite{johnston2014modeling,esposito2008o2,andrienko2017state} \\
 \hline
 X$_2^f$ & \(\mathrm{O}_2 + \mathrm{N} \rightarrow \mathrm{NO} + \mathrm{O}\) & - & - & - & & \cite{armenise2021n+,armenise2023low} \\
 \hline
 \textcolor{blue}{$^b$}X$_3^f$ & \(\mathrm{NO} + \mathrm{O}_2 \rightarrow \mathrm{NO}_2 + \mathrm{O}\) & \(1.80 \times 10^{-20}\) & 0.58 & 1.96 & 1 & \cite{atkinson1989evaluated} \\
 \hline
 \textcolor{blue}{$^c$}X$_4^f$ & \(\mathrm{N}_2 + \mathrm{M} \rightarrow \mathrm{N} + \mathrm{N} + \mathrm{M}\) & \(1.16 \times 10^{-8}\) & -1.6 & 9.75 & 1 & \cite{johnston2014modeling,esposito2006n} \\
 \hline
 X$_5^f$ & \(\mathrm{N}_2 + \mathrm{O} \rightarrow \mathrm{NO} + \mathrm{N}\) & - & - & - & & \cite{esposito2017reactive,armenise2023low} \\
 \hline
 X$_6^f$ & \(\mathrm{NO}_2 + \mathrm{NO}_2 \rightarrow \mathrm{NO} + \mathrm{NO} + \mathrm{O}_2\) & \(6.56 \times 10^{-18}\) & 0 & 1.20 & & \cite{konnov2009implementation} \\
 \hline
 \textcolor{blue}{$^d$}X$_7^f$ & \(\mathrm{NO} + \mathrm{M} \rightarrow \mathrm{N} + \mathrm{O} + \mathrm{M}\) & \(3.32 \times 10^{-15}\) & 0 & 6.50 & & \cite{johnston2014modeling} \\
 \hline
 X$_8^f$ & \(\mathrm{NO} + \mathrm{O} + \mathrm{M} \rightarrow \mathrm{NO}_2 + \mathrm{M}\) & \textcolor{blue}{$^e$}\(2.92 \times 10^{-40}\) & -1.41 & 0 & & \cite{yarwood1991direct} \\
 \hline
\end{tabular}

\footnotesize{\textcolor{blue}{$^a$} \textit{A} is multiplied by 5.0 for M = N atom. \myworries{The thermal rate coefficients for M = O$_2$ and O are derived from QCT data.}  }

\footnotesize{\textcolor{blue}{$^b$} The rate coefficients for the reaction NO+O$_2$ are calculated using detailed balance, based on the rate coefficients of the reaction NO$_2$+O. The parameters \textit{A}, \textit{b}, and \textit{E$_a$} are then fitted using the Arrhenius equation.}

\footnotesize{\textcolor{blue}{$^c$} \textit{A} is multiplied by 4.3 for M = O atom. \myworries{The thermal rate coefficients for M = N$_2$ and N are derived from QCT data.}}

\footnotesize{\textcolor{blue}{$^d$} \textit{A} is multiplied by 22 for atoms and NO, by 1.0 for other molecules, respectively.}

\footnotesize{\textcolor{blue}{$^e$} The unit for \secondround{three-body} reactions is m$^6$s$^{-1}$.}

\end{table}

\myworries{ \secondround{It is worth emphasizing that} QCT is primarily used in atom-molecule collisions \cite{esposito2008o2, esposito2006n}. For molecule-molecule collisions, as detailed by Andrienko \textit{et al.} \cite{andrienko2015high}, there \thirdround{are} insufficient QCT data available for modelling. Therefore, a semi-empirical scaling of QCT atom-molecule dissociation data is applied to the calculation of rate coefficients for molecule-molecule collisions. All QCT molecule-molecule collision data are sourced from the work of Esposito \textit{et al.} \cite{esposito2006n} and Andrienko \textit{et al.} \cite{andrienko2017state}.}

\myworries{The total non-thermal rate coefficients for all forward Zeldovich reactions (X$^f_2$, and X$^f_5$) are calculated using Eq.\ref{eq:MD}, with the Boltzmann factor determined by the vibrational temperature. \thirdround{Likewise}, the Boltzmann factor in Eq.\ref{eq:QCT} \secondround{is modified} to calculate the total non-thermal rate coefficients for the reactions O$_2$+O$_2$/O in reaction \(X_1^f\) and N$_2$+N$_2$/N in reaction \(X_4^f\).} Based on the comparison by Lino da Silva \textit{et al.}, the Fridman-Macheret method provides more accurate results for calculating non-thermal rate coefficients than other two-temperature methods \cite{da2007two}. Therefore, a generalized Fridman-Macheret method is employed to compute the total non-thermal rate coefficients for \secondround{reaction} X$^f_3$ and the remaining parts of X$^f_1$ and X$^f_4$, as given by:

\begin{equation}
\label{eq:knn}
k_{non}^{total}(T_g, T_v)=\ \frac{AT_g^be^\frac{{-E}_a}{k_BT_g}Z(T_g)\sum_{v}^{max}e^{(\frac{-E_{v}}{k_BT_v}+\frac{\alpha E_{v}^\ast}{k_BT_g})}}{Z(T_v)\sum_{v}^{max}e^\frac{-(E_{v}-\alpha E_{v}^\ast)}{k_BT_g}}
\end{equation}
where \(A\) is the pre-factor of the Arrhenius rate equation in Table~\ref{tab:reactions}, \(E_{v} \) [eV] is the vibrational energy at \(v^{th}\) vibrational level,  \(\alpha\) is the so-called \(\textquoteleft\)coefficient of vibrational energy utilisation', taking a value between 0 and 1 \cite{fridman2008plasma}, \(E_{v}^\ast\) is the effective vibrational energy of molecules, defined as the minimum value between \(E_{v}\) and  \(E_{a}\alpha^{-1}\), and \(Z\) is the vibrational partition function. More details on the procedure for calculating the non-thermal rate coefficients are reported by Shen \textit{et al.} \cite{shen2024two}.  In this model, the rate coefficients of the forward reactions X$_1^f$$-$X$_5^f$ can be enhanced by vibrational excitation. 

\begin{figure}[h t]
\centering
\includegraphics[width=1\linewidth]{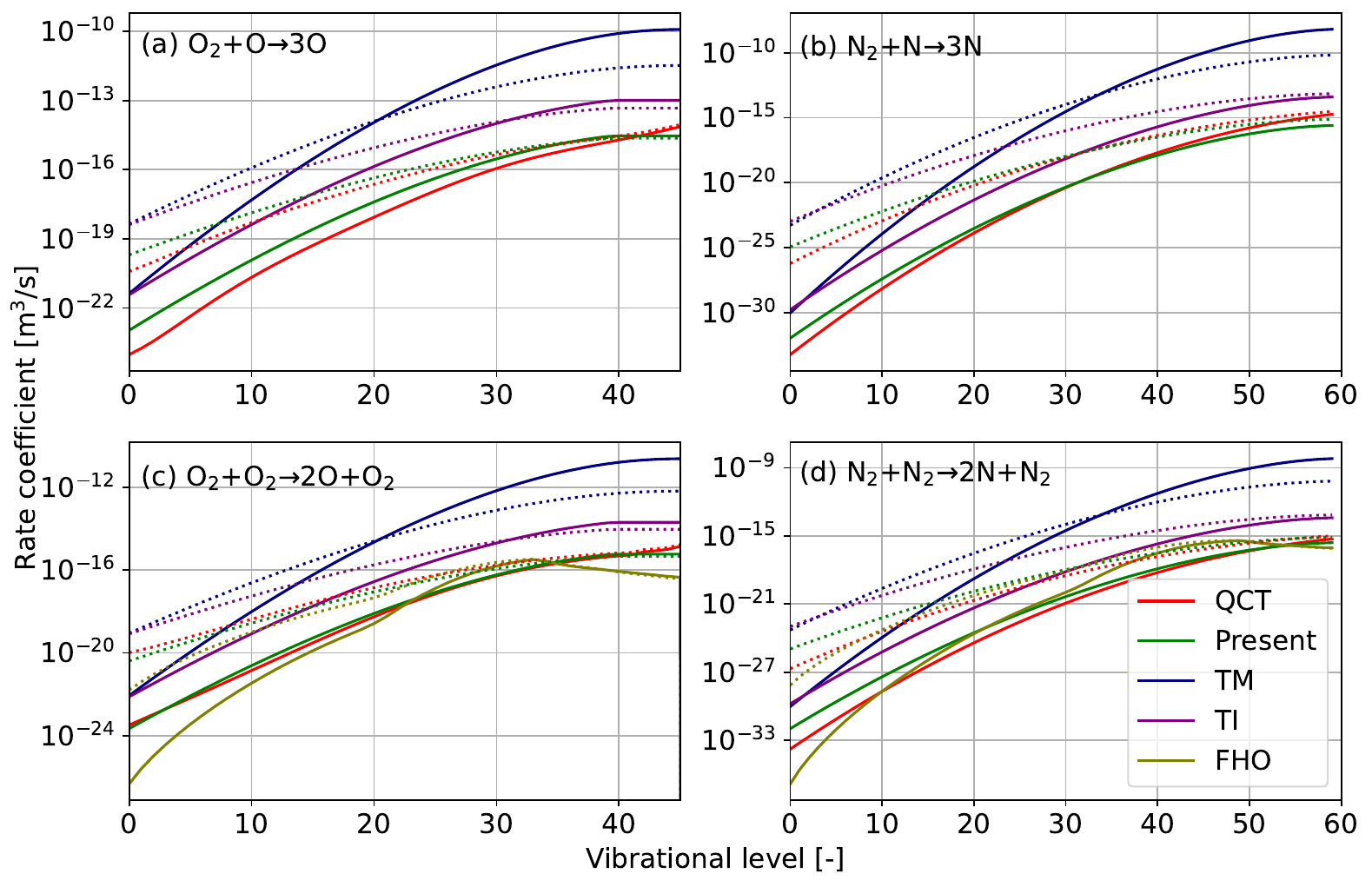}
\caption{ Comparison of V-D rate coefficients as a function of the vibrational level at different gas temperatures. Solid line: $T$ = 3000~K; dotted line: $T$ = 5000~K.  }
\label{vibraional_dissocaition_rate}
\end{figure}

A comparison of the \secondround{state-specific reaction} rate coefficients using the present method (Eq.\ref{eq:knn}), the theoretical-informational (TI) method \cite{berthelot2018pinpointing, fridman2008plasma}, the Forced Harmonic Oscillator (FHO) method \cite{adamovich1998vibrational,da2007state}, the Treanor-Marrone (TM) method ($U$ = 3$T$) \cite{capitelli2000rate} and the QCT method \cite{esposito2008o2, esposito2006n, andrienko2017state} is shown in Figure~\ref{vibraional_dissocaition_rate}.  It is clear that \secondround{the present} method agrees best with the QCT calculation, which provides the most accurate data within this temperature range. However, \secondround{the present} method requires \secondround{far} less computational cost \secondround{than} the QCT method. A more detailed discussion of the various rate coefficients and how they compare was reported by Esposito \cite{esposito2021reactive}.

\fourround{The cumulative reaction product over time is calculated by:}
\begin{equation}
\label{Eq.reaction production}
{\fourround{M(t)= S\int R_iudt}} 
\end{equation}
\fourround{where the integration is performed along the PFR by summarizing the contributions of individual CSTRs. \(S\) [m$^2$] is cross-section area, \(R_i\) [m$^{-3}$s$^{-1}$] is the rate of reaction \textit{i}, \(u\) [m$\cdot$s$^{-1}$] is the gas velocity, determined by local density and gas temperature:}
\begin{equation}
{\fourround{u(t) = \frac{\dot{m}}{S\rho(t)}=\frac{N(t)RT(t)}{PA} }}
\end{equation} 
\fourround{where \(\rho\) [kg$\cdot$m$^{-3}$] is mass density, \(P\) [Pa] is pressure,  \(N\) [mol$\cdot$m$^{-3}$] is molar density, \(\dot{m}\) is mass flow rate, fixed at 0.00021 kg s$^{-1}$.} 

The continuity equation for the different species is \cite{cantera_2}:

\begin{equation}
\fourround{{m\frac{dY_j}{dt} = \dot{m}(Y_{j,in}-Y_{j})+V\dot{\omega}_jW_j } }
\end{equation} 
\fiveround{where \(Y_{j,in}\) and \(Y_{j}\) are the mass fraction of species \textit{j} in the inflow and outflow, respectively.} \fourround{\(\dot{\omega}_j\) [mol$\cdot$m$^{-3}$s$^{-1}$] and \(W_j\) [kg$\cdot$mol$^{-1}$] are the molar production rate and molar mass of species \textit{j}, respectively. \(m\) [kg] represents mass.}

\subsection{Vibrational kinetics}

The vibrational excitation of N$_2$ and O$_2$ is recognized for its potential to lower the reaction barriers of the Zeldovich mechanism (reactions X$_2^f$ and \myworries{X$_5^f$} in Table \ref{tab:reactions}) \cite{kelly2021nitrogen, winter2021n2}. In this work, 60 vibrational levels of N$_2$ and 47 vibrational levels of O$_2$ are included, reported by Esposito \textit{et al.} \cite{esposito2017reactive, esposito2008o2} (Tables S1 and S2 in Supplementary material). The processes leading to vibrational excitation and de-excitation of molecular particles by heavy particle impact, encompassing Vibration-Vibration (V-V) and Vibration-Translation (V-T) energy relaxation, are presented in Table~\ref{tab:VVVT}. Many different methods have been developed to calculate the rate coefficients for V-V and V-T processes. The Schwartz–Slawsky–Herzfeld (SSH) method provided a theoretical framework for computing relaxation times in gases \cite{schwartz1952calculation}. However, its accuracy is limited to single-quantum transitions and \secondround{is insufficient for accurately calculating} V-V and V-T rates at high vibrational levels and gas temperatures \cite{guerra2019modelling, annuvsova2018kinetics}. To address these shortcomings, the FHO method, which was initially proposed by Kerner \cite{kerner1958note} and subsequently refined by Adamovich \textit{et al.} \cite{adamovich1998vibrational} and Lino da Silva \textit{et al.} \cite{da2007state}, incorporates the coupling of many vibrational states during a collision, \secondround{addressing} the limitations of the SSH method~\cite{adamovich1998vibrational, adamovich1995vibrational}. Unlike the SSH and FHO methods, which rely on the oscillator model, the QCT method offers a more comprehensive approach by considering realistic potential energy surfaces \cite{andrienko2018kinetic, armenise2023low}. It is important to note that while the QCT method \secondround{provides} superior accuracy, \secondround{its} prohibitive computational demands \secondround{restrict its applicability} \cite{da2009review}. To date, only a few databases for V-T relaxation \secondround{have been} calculated (\secondround{\textit{i.e.,}} O$_2$-O and N$_2$-N), and comprehensive databases for other vibrational exchange rates using the QCT method \secondround{are yet to be developed}.

\begin{table}[ht]
 \caption{ List of vibrational kinetics processes. v$_i$ and v$_n$ are less than 60. v$_j$ and v$_m$ are less than 46. Backward rate coefficients are calculated using the principle of detailed balance.}
 \label{tab:VVVT}
 \centering
 \begin{tabular}{ccc}

 \hline
  Type  &Process  & Ref  \\

 \hline
 V-T &N$_2$(v$_i$)+N$_2$ $\rightarrow$  N$_2$(v$_i$-\(\Delta v\))+N$_2$, 1\(\leq \Delta v \leq\) v$_i$      &  \cite{adamovich1998vibrational}\\
 \hline
 V-T &N$_2$(v$_i$)+O$_2$ $\rightarrow$  N$_2$(v$_i$-\(\Delta  v\))+O$_2$, 1\(\leq \Delta v \leq\) v$_i$     & \cite{adamovich1998vibrational}\\
 \hline
 V-T &N$_2$(v$_i$)+N $\rightarrow$  N$_2$(v$_i$-\(\Delta v\))+N, 1\(\leq \Delta v \leq\) min(v$_i$, 50)    & \cite{esposito2006n}\\
 \hline
 V-T &N$_2$(v$_i$)+O $\rightarrow$  N$_2$(v$_i$-1)+O      & \cite{popov2011fast}\\
 \hline
 V-T &O$_2$(v$_j$)+O $\rightarrow$  O$_2$(v$_j$-\(\Delta v\))+O , 1\(\leq \Delta v \leq\) min(v$_i$, 30)    & \cite{esposito2008o2}\\
 \hline
 V-T &O$_2$(v$_j$)+O$_2$ $\rightarrow$  O$_2$(v$_j$-\(\Delta v\))+O$_2$, 1\(\leq \Delta v \leq\) v$_j$     & \cite{adamovich1998vibrational} \\
 \hline
 V-T &O$_2$(v$_j$)+N$_2$ $\rightarrow$  O$_2$(v$_j$-\(\Delta v\))+N$_2$, 1\(\leq \Delta v \leq\) v$_j$     & \cite{adamovich1998vibrational} \\
 \hline

 V-V &O$_2$(v$_j$)+N$_2$(v$_i$-1) $\rightarrow$ O$_2$(v$_j$-1)+N$_2$(v$_i$)    & \cite{adamovich1998vibrational}\\
 \hline

 V-V &O$_2$(v$_j$)+O$_2$(v$_m$-1) $\rightarrow$ O$_2$(v$_j$-1)+O$_2$(v$_m$)    & \cite{adamovich1998vibrational}\\
 \hline
 
 V-V &N$_2$(v$_i$)+N$_2$(v$_n$-1) $\rightarrow$ N$_2$(v$_i$-1)+N$_2$(v$_n$)    & \cite{adamovich1998vibrational}\\
 \hline

 \hline

\end{tabular}
\end{table}

\begin{figure}[t]
\centering
\includegraphics[width=1\linewidth]{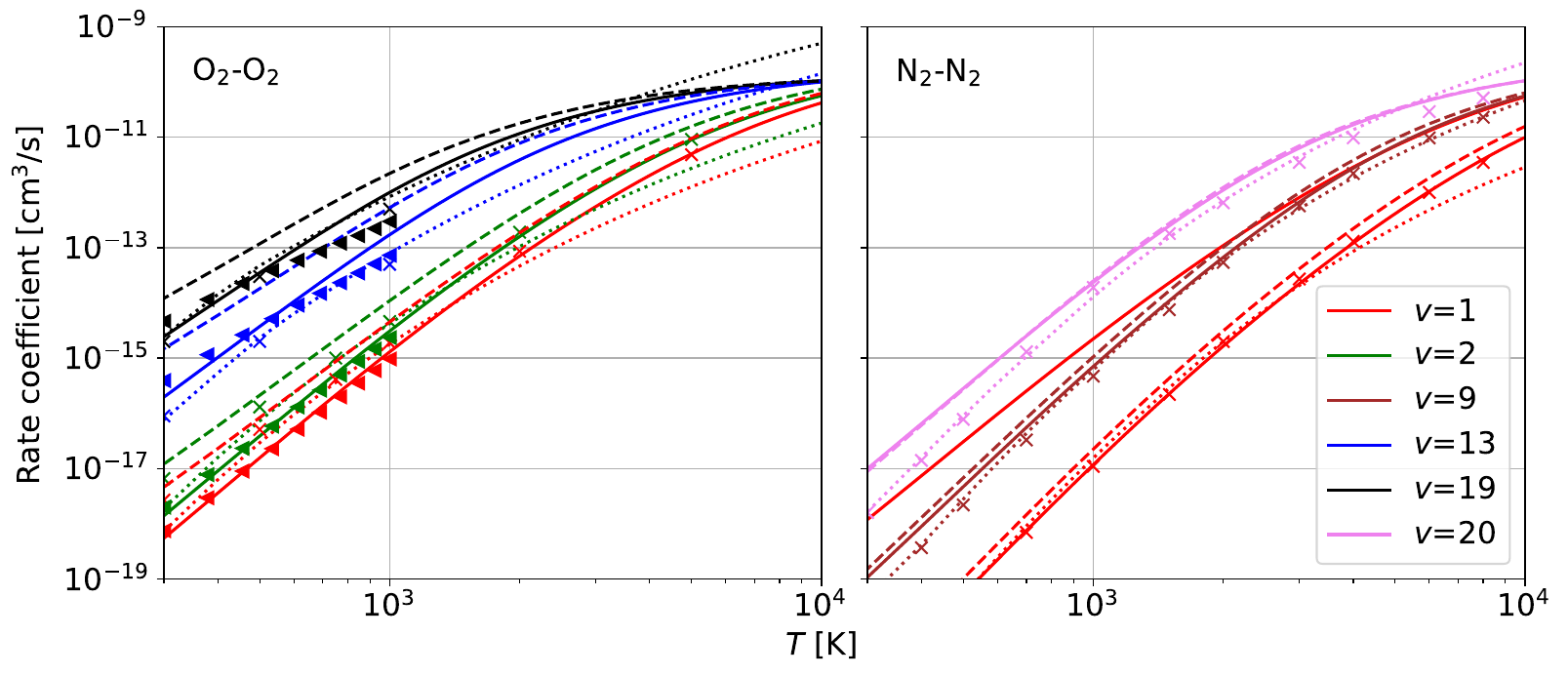}
\caption{Comparison of available single-quantum V-T rate coefficients calculated by different methods as a function of the gas temperature. Solid line: FHO calculations by the present work; dotted line: SSH calculations by Guerra  \textit{et al.} and Kozak  \textit{et al.} \cite{guerra2019modelling, kozak2014splitting}; dashed line: FHO calculations by Lino da Silva \textit{et al.} \cite{esther}; \(\times\): The \myworries{QC} calculations by Billing  \textit{et al.} \cite{billing1992vibrational, billing1979vv, billing1994vv}; \ding{115}: FHO calculations by Annušová  \textit{et al.}\cite{annuvsova2018kinetics}.}
\label{fig: VT_picture}
\end{figure}

A comparison of single-quantum deactivation V-T rate coefficients using different methods is shown in Figure \ref{fig: VT_picture}. The SSH calculations of V-T O$_2$-O$_2$ and N$_2$-N$_2$ processes \cite{guerra2019modelling,kozak2014splitting} exhibit better agreement with the \myworries{quantum-classical (QC)} calculations at low temperatures (below 1000 K). However, the SSH method fails to offer accurate results for the condition of interest in this work, since the gas temperature in the MW plasma typically exceeds several thousand Kelvin \cite{tatar2024analysis, altin2022energy}. Therefore, the FHO method \secondround{is applied} to compute the majority of V-T processes (involving O$_2$-O$_2$, O$_2$($v$)-N$_2$, N$_2$-N$_2$, and N$_2$($v$)-O$_2$), as well as all V-V processes considered \secondround{in this study}. All data \secondround{related to the FHO calculation}, except \secondround{for} the vibrational energy of N$_2$ and O$_2$ \secondround{at} each vibrational level, can be found in the work of Adamovich \textit{et al.} \cite{adamovich1998vibrational}. \secondround{A striking difference is observed between the} V-T deactivation rates of the first vibrational state for O$_2$ and N$_2$, particularly at temperatures below 1000 K (\secondround{as depicted} by the red solid lines in Figure \ref{fig: VT_picture}). \secondround{O$_2$ exhibits notably faster vibrational deactivation than N$_2$, primarily due to its smaller vibrational energy spacing in the molecular manifold.} The discrepancy diminishes at higher temperatures.

\begin{figure}[t]
\centering
\includegraphics[width=0.6\linewidth]{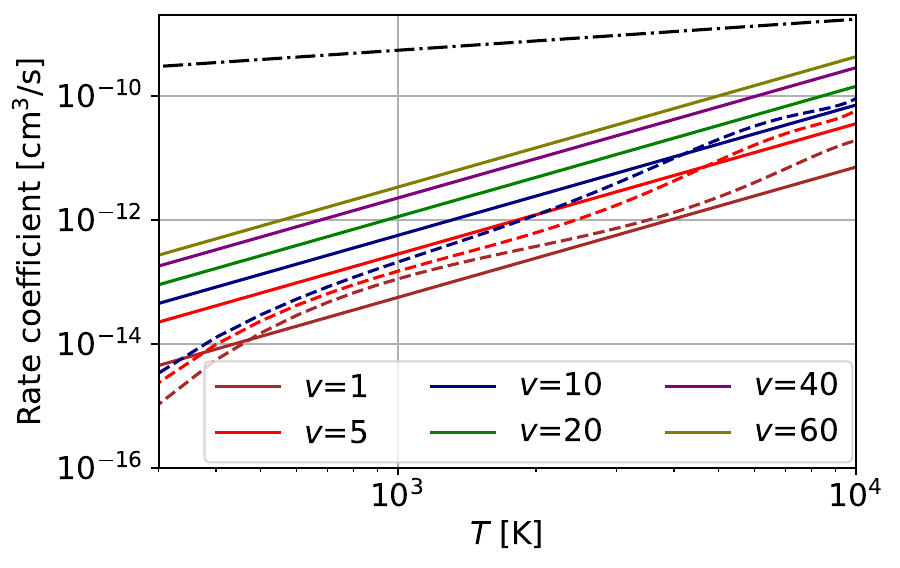}
\caption{Evolution of single-quantum V-T N$_2$-O ($v$\(\rightarrow\)$v$-1) process as a function of gas temperature. Solid line: (scaled) fit of the experimental data \cite{popov2011fast,guerra2019modelling}; dashed line: results from Hong \textit{et al.}\cite{hong2022vibrational}; dash-dotted line: the gas-kinetic collision rate\cite{da2007state}.}
\label{fig: V_T_N2_O}
\end{figure}

The V-T N$_2$-O process is considered to be one of the primary deactivation mechanisms, but the QCT results for the reaction rate are lower \secondround{than the experimental results} by orders of magnitude \cite{esposito2022relevance, hong2022vibrational}. Therefore, the V-T rate coefficients of the N$_2$-O relaxation from the first vibrational state to the ground state is taken from Popov \cite{popov2011fast}, which is fitted based on available experimental data \cite{breshears1968effect, eckstrom1973vibrational, mcneal1974temperature}. Given the exceptionally high probabilities of the V-T N$_2$-O process at elevated vibrational levels, the V-T rate coefficients at high levels are determined using the harmonic oscillator scaling law \cite{guerra1995non}. It means that only the single-quantum V-T N$_2$-O relaxation processes are considered in the model. The gas-kinetic collision rates are \secondround{introduced} to assess the reasonableness of the V-T rate coefficients under varying conditions. All V-T rate coefficients remain \secondround{lower than} the gas-kinetic rate, even at very high temperatures (up to 10$^4$ K in this study) and elevated vibrational levels, as shown in Figure \ref{fig: V_T_N2_O}. Additionally, despite some deviations observed at lower temperatures and higher quantum numbers, in general the rate results align well with those obtained using the mixed quantum-classical (MQC) \secondround{and Landau-Zener methods} by Hong \textit{et al.} \cite{hong2022vibrational}. \secondround{To enhance accuracy, more rate coefficients for V-T N$_2$-O processes are still needed, in particular to account for multi-quanta relaxation, using the MQC methods or other appropriate approaches.}

For the V-T O$_2$-O and N$_2$-N processes, the interpolated data of the QCT method from Esposito \textit{et al.} \secondround{is used directly} \cite{esposito2006n, esposito2008o2}.  \myworries{Given the relatively weak non-equilibrium conditions studied in this work, the coexistence of N atoms with O$_2$ is rare under various conditions. \secondround{Figures illustrating the time evolution of mole fractions in the quenching region at different initial temperatures and cooling rates can be found in Section \ref{sec:mole_fraction_quenching} and  Supplementary material. These figures also demonstrate the limited availability of N atoms coexisting with O$_2$ in the quenching region.} Consequently, the V-T O$_2$-N process can be safely neglected in the simulation, thereby reducing computational \secondround{costs}. }



\begin{figure}[t]
\centering
\includegraphics[width=0.6\linewidth]{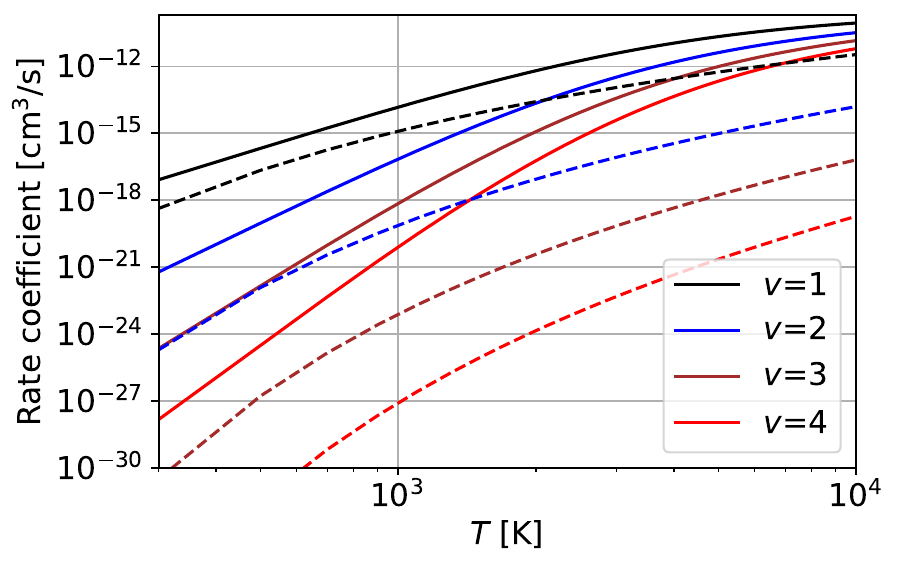}
\caption{Vibrational relaxation N$_2$+O$_2$ \thirdround{collisions} as a function of gas temperature. Solid line: V-T N$_2$-O$_2$ (5\(\rightarrow\)5-\(v\)) process; dashed line: V-V N$_2$-O$_2$ ((5,0)\(\rightarrow\)(5-\(v\),\(v\))) process.}
\label{fig: VVandVTrate}
\end{figure}

In addition to single-quantum transitions, exploring the necessity of multi-quantum transitions in V-V and V-T processes is crucial for saving computational resources while maintaining accuracy (Figure \ref{fig: VVandVTrate}). As the number of quanta increases, V-V rate coefficients exhibit a substantial decrease of nearly three orders of magnitude at different temperatures. Furthermore, it is worth noting that the computational cost required for the whole V-V process is significantly higher than that for the V-T process. For instance, \secondround{with the inclusion of} reverse relaxation reactions, the entire single-quantum V-V N$_2$-O$_2$ process encompasses 5104 reactions, whereas the entire single-quantum transition of the V-T N$_2$($v$)-O$_2$ process covers only 116 reactions. Consequently, to address computational constraints, this study \secondround{considers} solely single-quantum V-V energy transition. For the V-T energy transition, while the rate coefficients of single-quantum transition consistently remain the highest, the differences \secondround{in rate coefficients among various} multi-quantum transitions are \secondround{much} smaller than those observed in the V-V processes, especially at high temperatures. Furthermore, it is important to note that, compared with the single-quantum V-T process, more vibrational energy can be dissipated in a multi-quantum V-T transition. Consequently, these processes can still play a significant role in vibrational energy relaxation. \secondround{For instance}, Figure \ref{fig:S3} compares the energy transfer rates including V-T processes with different maximal numbers of quanta. The results emphasize the necessity of the multi-quantum V-T relaxation, especially at high temperatures. Therefore, nearly all the multi-quantum transitions in the V-T process are included in the model (Table~\ref{tab:VVVT}).

\subsection{Gas heating mechanisms}
The heat exchange with the surroundings is the primary cooling mechanism in the model. The power density loss by external cooling is defined as \cite{vermeiren2020plasma, pintassilgo2014study}:

\begin{equation}
\label{cooling equation}
{Q_{wall}=c\frac{8\lambda }{r^2}(T-T_w)} 
\end{equation}
where \(T_w\) is \secondround{the} wall temperature, fixed at 298.15~K, \(r\) is the reactor inner radius of the PFR (taken here as 13.5~mm), \(\lambda\) [W(m$\cdot$K)$^{-1}$] represents the thermal conductivity. However, due to the influence of gas flow turbulence, the cooling rate in reality can be \secondround{considerably} higher \cite{belov2018carbon}. Additionally, external factors such as using a nozzle \cite{van2022effusion, van2024effluent} or \secondround{a} water-cooled quenching rod \cite{kim2020carbon} can further enhance the cooling in the afterglow. Therefore, a factor \(c\) ranging from 10 to 1000 \secondround{is introduced} to artificially raise the cooling rate in the model, making it more representative of experiments \cite{vermeiren2020plasma}.

\myworries{Based on the work of Kustova \textit{et al.} \cite{kustova2006correct}, the vibrational thermal conductivity \secondround{ for species \textit{k}} is calculated by:}

\begin{equation}
{ \myworries{\lambda^{vib}_{k}= \frac{N_{k}\fourround{C_{vib}^{k}}}{\sum_j \frac{X_j}{D_{jk}}}}}
\label{vib_lamda}
\end{equation}
\myworries{where \(N_{k}\) [mol$\cdot$m$^{-3}$] is the molar density of species~\textit{k}, \(X_k\) is the molar fraction of species \textit{k}, \(D_{jk}\) [in m$^2$s$^{-1}$]  is the binary diffusion coefficient between species \(j\) and \(k\), which depends on temperature and is calculated based on the Lennard-Jones binary interaction potential \cite{synek2015interplay,vialetto2022charged,kee1986fortran}. \fourround{\(C_{vib}^ {k}\)} [J(mol\(\cdot\)K)$^{-1}$] is the vibrational molar heat capacity, which is equal to:}
\begin{equation}
{\fourround{C_{vib}^ {k}} =N_A\frac{d E^{k}_{ave} }{d T_v^{k}} } 
\end{equation}
where \(N_A\) is Avogadro's constant, and \(E^{k}_{ave}\) is \secondround{the} average vibrational energy \cite{andrienko2015high},  which is calculated by:
\begin{equation}
\label{Ev_equ}
{E^{k}_{ave}= \sum_v^{max}E_v^{k}f_v(T_v^{k}) } 
\end{equation}
where \(E_v^{k}\) is the vibrational energy of N$_2$ or O$_2$ \secondround{at} the vibrational level $v$, and \(f_v(T_v^{k})\) is the Boltzmann factor based on the vibrational temperature.

\myworries{A semi-empirical formula is used to calculate the thermal conductivity of the mixture \(\lambda_{mix}\) \cite{mathur1967thermal, kee2005chemically,vialetto2022charged}:}
\begin{equation}
{ \myworries{\lambda_{mix}= \frac{1}{2} \left( \sum X_k \lambda_k +  \frac{1}{\sum X_k / \lambda_k} \right) }}
\label{total_lamda}
\end{equation}
\myworries{where \(\lambda_k\) [W(m\(\cdot\)K)$^{-1}$], is the pure species heat conductivity, and the coefficients of the polynomials are tabulated in the NASA library \cite{mcbride1993coefficients}. In order to exclude the contribution of the vibrational modes of N$_2$ and O$_2$, the pure-species thermal conductivity of N$_2$ and O$_2$ in Eq.\ref{total_lamda} should be replaced by \(\lambda_{k}^{tr-ro}\), which is defined by Warnatz \cite{kee1986fortran}: }

\begin{equation}
{\myworries{\lambda_{k}^{tr-ro}= \lambda_{k}- N_{k}D_{kk} \fourround{C_{vib}^k }}} 
\label{lambda_tr}
\end{equation}
\myworries{where \(D_{kk}\) [in m$^2$s$^{-1}$] is the self-diffusion coefficient, which is fitted by NASA \cite{gupta1990review}.} \secondround{All values in Eq.\ref{lambda_tr} are relevant to \(T_g\).}  \fourround{Therefore, the power densities loss by external cooling relevant to vibrational and gas temperature are calculated by using different conductivities and temperatures based on Eq.\ref{vib_lamda} and Eq.\ref{lambda_tr}, with sharing the same value of \textit{c} in Eq.\ref{cooling equation}.}

\secondround{
Given that \(\lambda_{mix}\) depends on both the gas composition and gas temperature, the cooling rate is not constant within the quenching region. \thirdround{The cooling timescale is defined by the characteristic cooling time \(\tau\) [s], over which the difference between the gas and wall temperatures becomes $e$ times less than in the beginning. The average cooling rate $\dot{T}$ [K/s] over this time is:}
\begin{equation}
\label{charater_time}
{\dot{T}=\frac{T_g^0 (1-1/e)}{\tau} } 
\end{equation}
where \(T_g^0\) [K] represents the initial gas temperature. Since the power density loss (\(Q_{wall}\)) due to external cooling depends on the temperature difference between the gas and wall temperatures, the relaxation time decreases with rising initial gas temperature. For different values of \textit{c} (10, 100, and 1000), the average cooling rates are on the order of magnitude of 10$^6$ K·s$^{-1}$, 10$^7$ K·s$^{-1}$, and 10$^8$ K·s$^{-1}$, respectively. For instance, as the initial gas temperature is 3000 K, the average cooling rates are 2.2\(\times\)10$^6$~K·s$^{-1}$, 2.1\(\times\)10$^7$~K·s$^{-1}$, and 1.8\(\times\)10$^8$~K·s$^{-1}$ under the conditions of c=10, 100, and 1000, \thirdround{respectively}.}

In the non-thermal quenching process, the difference between the gas temperature and vibrational temperatures necessitates consideration of \secondround{both} V-V and V-T processes. The processes involving electronically excited states are assumed to contribute \secondround{negligibly} to the vibrational power density in the quenching process. The vibrational temperature as a function of position is calculated by \fourround{\cite{cantera}}: 

\begin{eqnarray}
\label{cp_vib}
N C_{vib}^{N_2(O_2)} \frac{d T_v^{N_2(O_2)}}{d t} &=&
\frac{\dot{m} Y_{N_2(O_2),in}}{V} \left(h_{vib,in}^{N_2(O_2)} - h_{vib}^{N_2(O_2)} \right) + \sum Q_{chem-vib}^{N_2(O_2)} \nonumber \\
&&
- \sum Q_{VV}^{N_2(O_2)+M}
- \sum Q_{VT}^{N_2(O_2)+M}
- Q_{wall}^{vib}
\end{eqnarray}
where \fourround{\(h_{vib,in}^{N_2(O_2)}\) [J\(\cdot\)kg$^{-1}$] and \(h_{vib}^{N_2(O_2)}\) [J\(\cdot\)kg$^{-1}$] are the vibrational energy of inflow and outflow, respectively.} \(Q_{VT}^{N_2(O_2)+M}\) and \(Q_{VV}^{N_2(O_2)+M}\) [W$\cdot$m$^{-3}$] are the vibrational power density exchange by related V-T and V-V processes of N$_2$ and O$_2$, separately (Table~\ref{tab:VVVT}). \myworries{\(Q_{wall}^{vib} \) represents the vibrational power density loss by conduction, based on Eq.\ref{cooling equation} and Eq.\ref{vib_lamda}.} \secondround{However, it should be stressed that this term could be overestimated \thirdround{at low temperatures} due to slow surface vibrational relaxation \cite{black1974measurements, marinov2012surface}.} \secondround{\(\sum Q_{chem-vib}^{N_2(O_2)}\)} [W$\cdot$m$^{-3}$] represents the total vibrational power density resulting from related chemical reactions for either O$_2$ or N$_2$, which contributes to raising or lowering the vibrational temperatures (Table \ref{tab:reactions}). It is worth mentioning that  vibrational energy can be both supplied by exothermic chemical reactions and consumed by endothermic chemical reactions. Therefore, it is essential to calculate how the released or consumed energy is distributed between vibrational and translational-rotational degrees of freedom. \(\sum Q_{chem-vib}^{N_2}\) and  \(\sum Q_{chem-vib}^{O_2}\) can be expressed as:
\begin{equation}
\label{Ev_equ}
{\sum Q_{chem-vib}^{O_2}= E_b^{O_2}\sum_{m=1,2,3}(R_m^b\gamma_{rec}^{m} - R_m^f\gamma_{dis}^{m})- E_{ave}^{O_2}\sum_{m=1,2,3}(R_m^b-R_m^f)} 
\end{equation}
\begin{equation}
\label{Ev_equ}
{\sum Q_{chem-vib}^{N_2}= E_b^{N_2}\sum_{m=4,5}(R_m^b\gamma_{rec}^{m} - R_m^f\gamma_{dis}^{m})- E_{ave}^{N_2}\sum_{m=4,5}(R_m^b-R_m^f)} 
\end{equation}
where \(E_b^{N_2}\) and \(E_b^{O_2}\) are the bond energies (\textit{i.e.,} potential energy of infinitely separated atoms relative to the ground vibrational state) of N$_2$ (9.75 eV) and O$_2$ (5.11 eV), \(m\) represent the number of reactions in Table \ref{tab:reactions}.

\begin{figure}[t]
\centering
\includegraphics[width=0.6\linewidth]{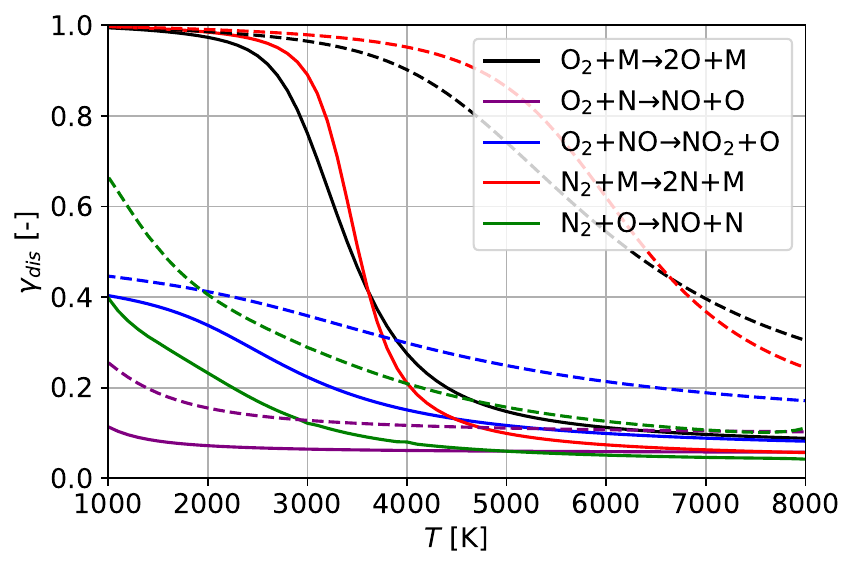}
\caption{The fractions of the bond energy provided by vibrational energy in dissociation (\(\gamma_{dis}\)) for various reactions as a function of gas temperature at different vibrational temperatures. Solid line: $T_v$ = 3000 K; dashed line: $T_v$ = 5000 K. The \(\gamma_{dis}\) of O$_2$+M is determined as the average of the values for the reactions O$_2$+O and O$_2$+O$_2$, derived from QCT data provided by Esposito \textit{et al.} \cite{esposito2008o2} and Andrienko \textit{et al.} \cite{andrienko2017state}, respectively. The \(\gamma_{dis}\) of N$_2$+M is calculated as the average of the values for the reactions N$_2$+N and N$_2$+N$_2$, based on QCT data from Esposito \textit{et al.} \cite{esposito2006n}. The \(\gamma_{dis}\) of Zeldovich reactions N$_2$+O and O$_2$+N are derived using QCT data from Esposito \textit{et al.} \cite{esposito2017reactive} and Armenise \textit{et al.} \cite{armenise2021n+,armenise2023low}. Additionally, the required non-thermal rate coefficients for the reaction NO+O$_2$ are calculated using the generalized Fridman-Macheret method \cite{shen2024two}.   }
\label{fig: gamma_dis}
\end{figure}

The values of $\gamma_{dis}$ (\textit{i.e.,} the fraction of the bond energy provided by the vibrational energy in dissociation) can be determined based on the non-thermal rate coefficients:
\begin{equation}
{\gamma_{dis}= \frac{\sum_i^{max} f_i( T_v^{N_2(O_2)}) \secondround{k_{i}} E_i^{N_2(O_2)}}{E_b^{O_2(N_2)} \sum_j^{max} f_j( T_v^{N_2(O_2)}) \secondround{k_{j}} }} 
\end{equation}
where \(\secondround{k_{i}} \) [m$^3$s$^{-1}$] is the \secondround{state-specific} rate coefficient \secondround{at the} \(i^{th}\) vibrational state. Due to the Boltzmann factor, $\gamma_{dis}$ is affected not only by the gas temperature, but also by the vibrational temperature. Varied $\gamma_{dis}$ values \secondround{as functions} of the gas temperature with different vibrational temperatures are shown in Figure \ref{fig: gamma_dis}. Due to the absence of complete QCT databases for the vibration-dissociation rate coefficients of N$_2$+O$_2$ and O$_2$+N$_2$ at different vibrational levels, the $\gamma_{dis}$ of O$_2$+M and N$_2$+M are approximated using the average values of the reactions O$_2$+O$_2$(O) and N$_2$+N$_2$(N), separately. For the dissociation reactions O$_2$+M and N$_2$+M, nearly all the chemical energy is derived from vibrational energy when the vibrational temperature significantly exceeds the gas temperature. As the gas temperature rises above the vibrational temperature, both values of $\gamma_{dis}$ decrease markedly. For the other three reactions (X$_2^f$, X$_3^f$, and X$_5^f$), the decrease in $\gamma_{dis}$ with increasing gas temperature occurs when the gas temperature is considerably lower than the vibrational temperature. This difference in behaviour compared to the dissociation reactions can be explained by the activation barriers being lower than the bond energies.

The fraction of the bond energy converted to the vibrational energy in recombination ($\gamma_{rec}$) can be defined as:
\begin{equation}
\gamma_{\mathrm{rec}} = 
\frac{
\sum_{i}^{\mathrm{max}} \mathrel{\mathop{k_i}\limits^{\leftarrow}} \, E_i^{N_2(O_2)}
}{
E_b \sum_{j}^{\mathrm{max}} \mathrel{\mathop{k_j}\limits^{\leftarrow}}
}
\label{gamma_rec}
\end{equation}
\noindent where \( \mathrel{\mathop{k_i}\limits^{\leftarrow}} \) [m$^3$s$^{-1}$] is the backward state-specific rate coefficient at the \(i^{th}\) vibrational state, calculated using the detailed balance principle:

\begin{equation}
\mathrel{\mathop{k_i}\limits^{\leftarrow}} = \frac{f_i(T_g) \, k_i}{K_{\mathrm{eq}}}
\end{equation}

\begin{figure}[ht]
\centering
\includegraphics[width=0.6\linewidth]{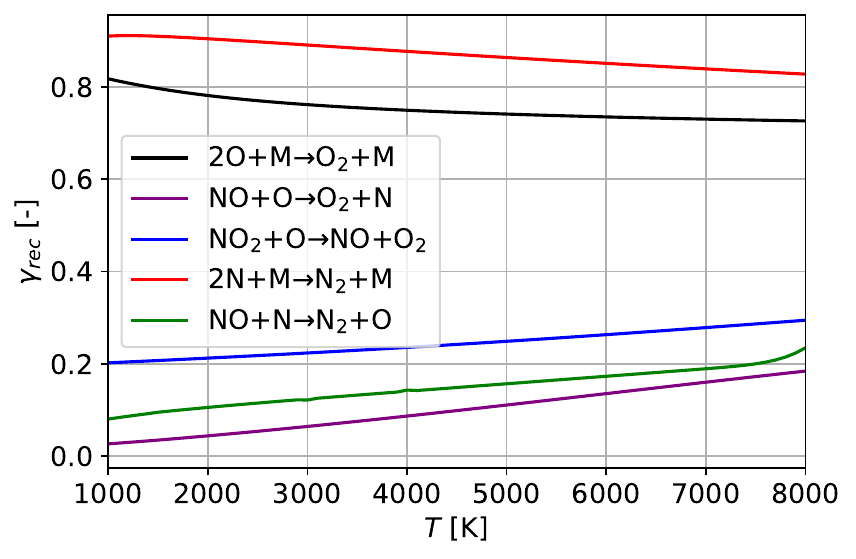}
\caption{The fraction of the bond energy converted to the vibrational energy in recombination (\(\gamma_{rec}\)) for various reactions as a function of gas temperature.}
\label{fig: gamma_rec}
\end{figure}

Therefore, Eq.\ref{gamma_rec} can be rewritten by:
\begin{equation}
{\gamma_{rec}= \frac{\sum_i^{max}  f_i(T_g) \secondround{k_i}  E_{i}^{N_2(O_2)}}{E_b \sum_{j}^{max} f_j(T_g) \secondround{k_j} } } 
\end{equation}
where \(\gamma_{rec}\) is influenced only by the gas temperature, and the value is equal to that of \(\gamma_{dis}\) under \secondround{the} thermal state (\(i.e.,\) \(T_g = T_v^{N_2(O_2)}\)). Consequently, $\gamma_{rec}$ has the same value under both thermal and non-thermal conditions, provided the gas temperature is unchanged. As the gas temperature increases, \secondround{reactions X$_b^1$ and X$_b^4$ show a slight decline in $\gamma_{rec}$, whereas the other reactions display an opposite trend,} as illustrated in Figure \ref{fig: gamma_rec}.

Since the calculation of \secondround{the} V-T power density shares the same equation for different V-T processes, \secondround{\(Q_{VT}^{O_2-O_2}\) is considered as} an example, which is equal to \cite{pintassilgo2014study,pintassilgo2016power}:


\begin{equation}
{Q_{VT}^{O_2-O_2}= n_{O_2}^2  \left\{ \sum_{r}f_r(T_v^{O_2}) \left( \sum_{w}P_{r,w}^{O_2-O_2} \Delta E^{O_2}_{r,w}   \right)   \right\}    } 
\end{equation}
where \(T_v^{O_2}\) [K] is the vibrational temperature of O$_2$, \(P_{r,w}^{O_2-O_2}\) [m$^{3}$s$^{-1}$] and \(\Delta E^{O_2}_{r,w}\) are the rate coefficients and \secondround{the} vibrational energy difference of \secondround{the} V-T O$_2$-O$_2$ process from \(r\)th \thirdround{vibrational} level to \(w\)th level, respectively.

Analogously, \(Q_{VV}\) [W$\cdot$m$^{-3}$] is the power density loss via the V-V process. Here, the vibrational power exchange of N$_2$ via the V-V N$_2$-O$_2$ process \secondround{is considered as an example, defined by} \cite{pintassilgo2014study, pintassilgo2016power}:
\begin{eqnarray}
Q_{VV(N_2)}^{N_2-O_2} &=&  n_{O_2}n_{N_2} \Bigg\{ \sum_{i=0}^{\mathrm{max}-1} f_i(T_v^{O_2}) \left( \sum_{n=1}^{\mathrm{max}} P_{N_2 (n,n-1)}^{O_2 (i,i+1)} f_n(T_v^{N_2}) \Delta E_{n,n-1}^{N_2} \right) \nonumber \\
&& - \sum_{n=1}^{\mathrm{max}} f_n(T_v^{N_2}) \left( \sum_{i=0}^{\mathrm{max}-1} P_{N_2 (n-1,n)}^{O_2 (i+1,i)} f_i(T_v^{O_2}) \Delta E_{n+1,n}^{N_2} \right) \Bigg\}
\end{eqnarray}
where \(T_v^{N_2}\) [K] is the vibrational temperature of N$_2$, \(P_{N_2  (n,n-1)}^{O_2 ( i,i+1)}\) and \(P^{N_2 (n-1,n)}_{O_2  (i+1,i)}\) [m$^{3}$s$^{-1}$] are the corresponding forward and backward rate coefficients of the V-V N$_2$-O$_2$ process, and both \(\Delta E_{ n,n-1}^{N_2}\) and \(\Delta E_{n+1,n}^{N_2}\) are vibrational \thirdround{energy} differences. It should be noted that due to the vibrational \thirdround{requency} difference between N$_2$ and O$_2$, a part of the vibrational energy of N$_2$ is lost \secondround{to} gas heating via the V-V process. Additionally, due to unharmonicity, the varying energy gaps between adjacent vibrational levels also cause the corresponding V-V relaxation processes (\(i.e.,\) N$_2$-N$_2$ and O$_2$-O$_2$) to release part of the energy as direct gas heating.

\fourround{The cumulative power loss of the V-V or V-T relaxation reaction \textit{i} over time can be calculated by:}
\begin{equation}
\label{eq:vib reaction production}
{ \fourround{C_i(t)=A\int Q_{VT/VV}^i udt} } 
\end{equation}  
\fourround{where the integration is performed along the PFR by summarizing the contribution of individual CSTRs.}

Because the translational-rotational equilibrium occurs almost instantly, translational and rotational modes are grouped, governed by \secondround{the} gas temperature (\textit{i.e.,} $T_g$ = $T_{tr}$ = $T_{rot}$) \cite{fridman2008plasma}. The calculation of the gas temperature is influenced by multiple factors, including chemical processes, conduction, V-T, and V-V relaxation processes. These factors collectively contribute to the dynamic evolution of \secondround{the} gas temperature during the quenching process \fourround{\cite{cantera}}:
\begin{eqnarray}
N C_P^{\mathrm{mix}}(T_g) \frac{d T_g}{dt} &=& \frac{\dot{m}}{V} \left( h_{\mathrm{in}} - \sum_j h_j Y_{j,\mathrm{in}} \right) + \sum Q_{\mathrm{chem}}^{\mathrm{heat}} \nonumber \\
&& - Q_{\mathrm{wall}}^{\mathrm{mix}} + \sum Q_{VV}^{\mathrm{heat}} + \sum Q_{VT}
\end{eqnarray}
where \fourround{\(h_{in}\) [J\(\cdot\)kg$^{-1}$)] and \(h_{j}\) [J\(\cdot\)kg$^{-1}$)] represent the total species enthalpy of the inflow (excluding vibrational modes) and the species enthalpy of component \(j\) in the outflow (also excluding vibrational modes), respectively.} \(\sum Q_{VV}^{heat}\) and \(\sum Q_{VT}\) are the sum of vibrational energy used for gas heating via all V-V and V-T relaxation processes, \thirdround{respectively}. \secondround{\(Q_{wall}^{mix}\) represents the power density loss of the gas mixture due to conduction, excluding contributions from the vibrational modes of N$_2$ and O$_2$,} \(C_P^{mix}\) [J(mol\(\cdot\)K)$^{-1}$] is the molar heat capacity of the gas mixture \thirdround{at constant pressure}, excluding the vibrational modes of N$_2$ and O$_2$,  which is calculated by:

\begin{equation}
{  C_P^{mix}(T_g)= \sum_{i\neq N_2,O_2}X_i C_p^i(T_g) + X_{N_2} C_{p(ro-tr)}^{N_2}(T_g) +   X_{O_2} C_{p(ro-tr)}^{O_2}(T_g)   } 
\end{equation}  
where \(C_p^i(T_g)\) [J(mol\(\cdot\)K)$^{-1}$] and \(X_i\) [-] are the molar heat capacity and molar fraction of \secondround{all} species except \secondround{for} N$_2$ and O$_2$. For N$_2$ and O$_2$, the molar heat capacity [J(mol\(\cdot\)K)$^{-1}$] of \secondround{the} rotational-translational mode is independent of temperature \cite{gupta1990review}. For the diatomic molecules N$_2$ and O$_2$, the values \secondround{of the molar heat capacity} are both equal to 3.5$R$ (\secondround{where 1.5$R$ comes from the translational component, 1$R$ accounts for gas expansion, and 1$R$ corresponds to the rotational component of diatomic species} \cite{gupta1990review}). Finally, \(Q_{chem}^{heat}\) [W$\cdot$m$^{-3}$] is the power density of \secondround{the} chemical energy converted for gas heating directly:

\begin{equation}
\sum Q_{chem}^{heat}=\sum Q_{chem} - \myworries{\sum Q_{chem-vib}^{N_2}- \sum Q_{chem-vib}^{O_2}}
\end{equation}
\fourround{where $\sum Q_{chem}$ represents the total power density released by all chemical reactions, calculated as follows:}
\begin{equation}
{\fourround{\sum Q_{chem}= \sum_i R_i \Delta H_i}} 
\end{equation}
\fourround{where \(\Delta H_i\) [J] is enthalpy change between products and reactants, where the reaction enthalpies include the translational and rotational enthalpy of the species evaluated at the gas temperature, and the vibrational enthalpy of the species evaluated at the vibrational temperature (\textit{i.e.} the average vibrational energy).}

\section{Results and discussion}

\subsection{Energy transfer and NO$_x$ production in the plasma region}

\begin{figure}[ht]
\centering
\includegraphics[width=0.6\linewidth]{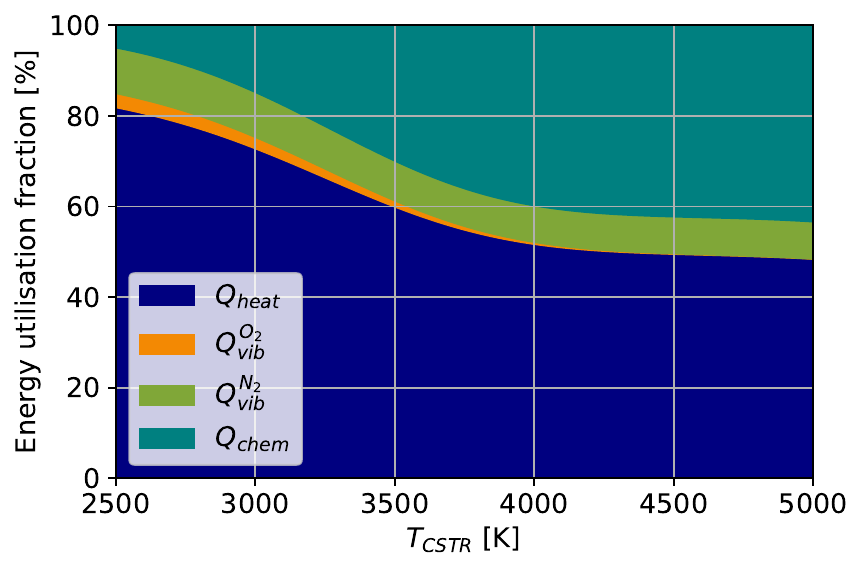}
\caption{ The energy utilisation fractions as a function of $T_{CSTR}$ in the CSTR model at atmospheric pressure. $Q_{heat}$ is the energy used for gas heating; $Q_{chem}$ is \secondround{the} chemical energy; $Q_{vib}^{O_2}$ and $Q_{vib}^{N_2}$ are the vibrational energy of O$_2$ and N$_2$, respectively.}
\label{energy utilisation in hot region}
\end{figure}

\begin{figure}[ht]
\centering
\includegraphics[width=0.6\linewidth]{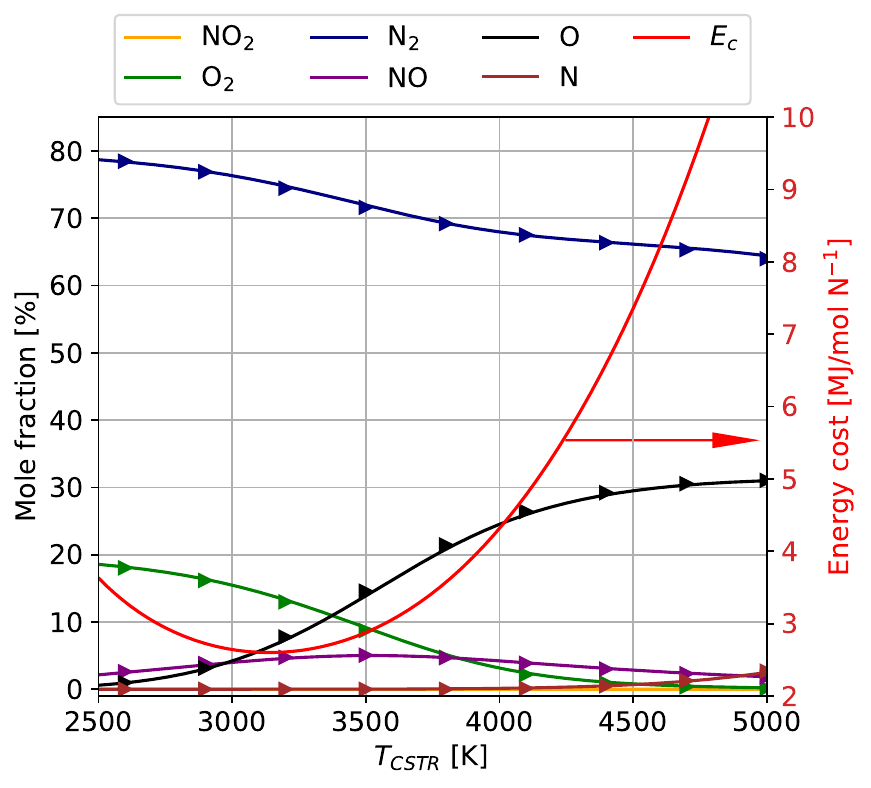}
\caption{The mole fractions and energy cost as a function of the temperature in the CSTR under atmospheric pressure. \secondround{Triangle symbols} represent thermodynamic equilibrium compositions calculated by D'Angola \textit{et al.} \cite{d2008thermodynamic}.}
\label{Mole fraction and energy cost in hot zone}
\end{figure}

As the initial air goes through the CSTR, which \secondround{is considered to be} in thermal equilibrium, a part of the heating power (in reality, provided by the MW radiation) is converted to chemical energy via chemical reactions, and heating is distributed uniformly over all degrees of freedom (Figure \ref{energy utilisation in hot region}). \secondround{As \(T_{CSTR}\) increases, an increasing fraction of O$_2$ dissociates into O atoms, resulting in a reduction of the vibrational energy of O$_2$.} With the assistance of active O atoms, the N$_2$ triple bond (9.75~eV) can be efficiently broken with an activation energy of 3.28~eV (reaction X$_5^f$). N atoms, produced in the above reaction, can subsequently react with O$_2$ to form NO (reaction X$_2^f$). Additionally, this reaction generates atomic oxygen, thus closing the chain of the Zeldovich mechanism~\cite{fridman2008plasma}. Figure \ref{Mole fraction and energy cost in hot zone} illustrates the mole fractions of species and energy cost \secondround{in the CSTR} as a function of $T_{CSTR}$. The energy cost of the CSTR is calculated under the absolute quenching condition, \textit{i.e.,} all the NO$_x$ production in the CSTR can be retained during the quenching. Under this quenching condition, the optimal energy cost in the CSTR model is 2.6~MJ/mol N$^{-1}$ at 3110 K  (Figure \ref{Mole fraction and energy cost in hot zone}). The molar fraction results show excellent agreement with the thermodynamic equilibrium compositions calculated by D’Angola~\textit{et~al.} \cite{d2008thermodynamic}. One of the major challenges in the plasma nitrogen fixation process is its high energy cost \cite{liu2024plasma}: even under optimal conditions, only 18\% of the total energy is utilized for dissociation \secondround{reactions} (\secondround{including those that are ineffective for synthesis}), and the remainder is wasted on gas heating (Figure \ref{energy utilisation in hot region}). Therefore, leveraging preheating by heat recovery and minimizing energy waste on gas heating are crucial steps to improve future performance \cite{vermeiren2020plasma, sarafraz2023conversion, patil2018plasma}.

Once NO is formed, it can be further oxidized to NO$_2$. However, nearly all the NO$_2$ produced by reaction X$_3^{net}$ is consumed by reaction X$_8^{net}$. Therefore, the selectivity of NO is much higher than that of NO$_2$ in the plasma region. With temperature increasing, although high temperatures can break more N-N chemical bonds, NO is already destructed, which limits the energy efficiency for NO production at high temperatures in the CSTR. Additionally, limited O$_2$ \secondround{availability} at high temperatures, due to reaction X$_1^f$, constrains NO production by reaction X$_2^{f}$. As a result, the reaction X$_5^{f}$ plays a dominant role in NO production at high temperatures. Above 4500~K, due to the depletion of O$_2$, the vibrational energy of O$_2$ decreases to zero (Figure \ref{energy utilisation in hot region}). Although the amount of N$_2$ also decreases with \secondround{higher} temperature (Figure \ref{Mole fraction and energy cost in hot zone}), a higher $T_{v}^{N_2}$ leads to the fraction of the vibrational energy of N$_2$ remaining stable.

\subsection{Energy redistribution and chemical processes in the quenching process}
\label{sec:mole_fraction_quenching}

\begin{figure}[ht]
\centering
\includegraphics[width=1\linewidth]{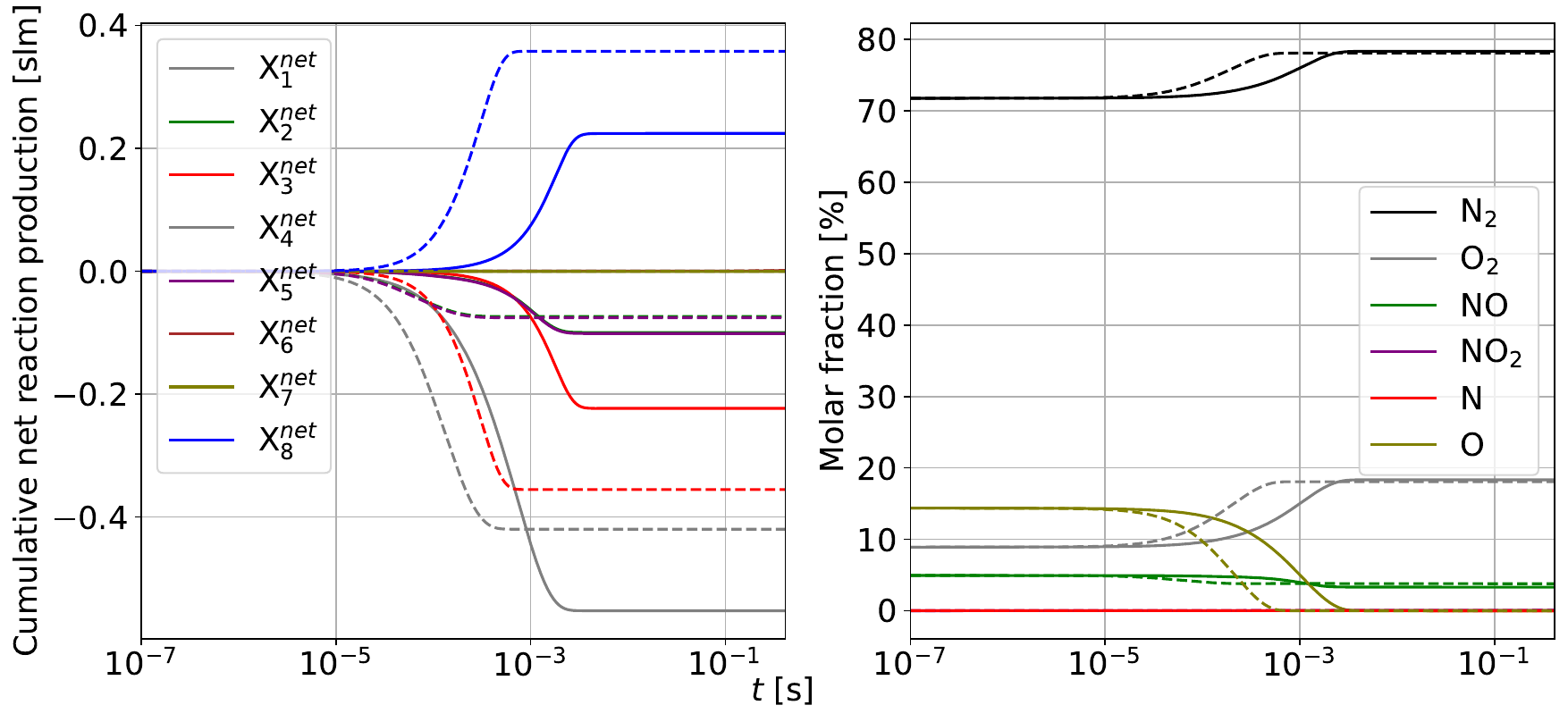}
\caption{\fourround{The cumulative net reaction product over different time intervals (based on Eq.\ref{Eq.reaction production})} and mole fractions as a function of the time in the quenching region with $T_{CSTR}$ = 3500~K at different cooling rates. Solid: $c$ = 10; dash: $c$ = 100. X$^{net}$ \fourround{represents that both forward and backward reactions are taken into account.}}
\label{reaction production(quenching)}
\end{figure}

With conduction to the wall, the gas temperature changes over time, triggering various reaction processes. \secondround{To illustrate}, the temporal evolution of the net production of various reactions and mole fractions of different species in the quenching process with $T_{CSTR}$ = 3500~K and at different cooling rates \secondround{is} shown in Figure \ref{reaction production(quenching)}. With the low cooling rate ($c$ = 10), \myworries{66.9\%} of O recombines to form O$_2$, and \myworries{13.6\%} reacts with NO to form NO$_2$. However, given that the reaction rate of X$_3^b$ is faster than that of reaction X$_8^f$, nearly all \secondround{of the} NO$_2$ reacts with another \myworries{13.6\%} of O atoms to form NO and O$_2$, \secondround{leading} to negligible net NO$_2$ formation until 0.1 s. A more detailed analysis of NO$_2$ formation is \secondround{provided in} the last section. The remaining (\myworries{5.9\%}) O atoms trigger the reverse Zeldovich mechanism (reactions X$_2^b$ and X$_5^b$), both of which reduce NO production. At a higher cooling rate ($c$ = 100), over \myworries{95.5\%} of O atoms \secondround{either recombine to form O$_2$ or react with NO and its subsequent products (\textit{i.e.,} NO$_2$) to generate O$_2$, effectively reducing NO loss via} the reverse Zeldovich mechanism. Likewise, the results for higher $T_{CSTR}$ (4500 K) can be found in Supplementary material.


\begin{figure}[ht]
\centering
\includegraphics[width=0.6\linewidth]{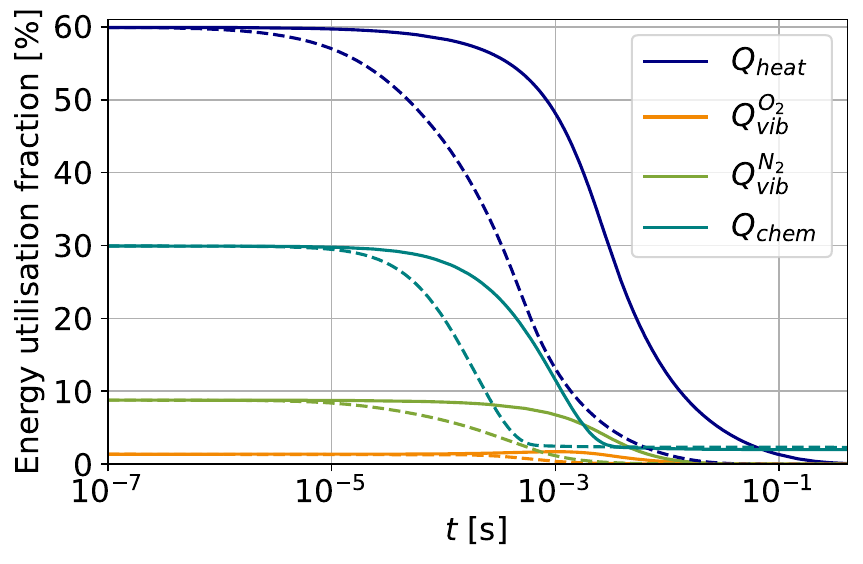}
\caption{The energy utilisation as a function of time with $T_{CSTR}$ = 3500~K at different cooling rates. Solid line: $c$ = 10; dashed line: $c$ = 100;}
\label{energy utilization in quenching process_3500K_950mbar_10}
\end{figure}

Figure \ref{energy utilization in quenching process_3500K_950mbar_10} depicts the temporal evolution of \secondround{the} energy utilisation fractions in the quenching process with $T_{CSTR}$ = 3500~K at different cooling rates. \secondround{At the start of the quenching process, which corresponds to the end of the CSTR}, nearly 59\% of the total energy is used for gas heating. Furthermore, a \secondround{small amount of} O$_2$ and most of \secondround{the} N$_2$ are still undissociated at the beginning of the quenching process (Figure \ref{Mole fraction and energy cost in hot zone}), leading to 1\% and 8\% of the total energy still stored as vibrational energy in O$_2$ and N$_2$, respectively. Due to the shorter \thirdround{characteristic cooling time (Figure \ref{temperature_differnet_cooling_rate})} compared to vibrational\thirdround{-translational} relaxation ($\tau_{vib}$) and the characteristic time of chemical reactions ($\tau_{chem}$), especially at high cooling rates, the translational-rotational energy relaxes faster than other forms of energy \cite{kosareva2022hybrid, kosareva2021four}. Subsequently, more and more O atoms recombine (Figure~\ref{reaction production(quenching)}), resulting in a portion of \secondround{the} recombination energy being converted to vibrational energy or heating the gas directly (Figure \ref{energy utilization in quenching process_3500K_950mbar_10}). Despite $\tau_{vib}$ being similar to $\tau_{chem}$, this portion of chemical energy can \secondround{keep} the vibrational energy of O$_2$ unchanged or even slightly increase \secondround{it before} 0.1~ms.  Because less chemical energy can convert to vibrational energy stored in N$_2$ than in O$_2$, the vibrational energy of N$_2$ decreases faster during the quenching process. Eventually, only \myworries{1.9\% and 2.2\%} of the total energy remains stored in the form of NO under the conditions of $c$ = 10 and 100, respectively, which represents the energy efficiency for NO$_x$ formation, while the rest of the energy is lost by heat transfer to the wall.

\subsection{Analysis of the loss processes for vibrational energy}
\label{section: Analysis of the loss processes for vibrational energy}

\begin{figure}[ht]
\centering
\includegraphics[width=1\linewidth]{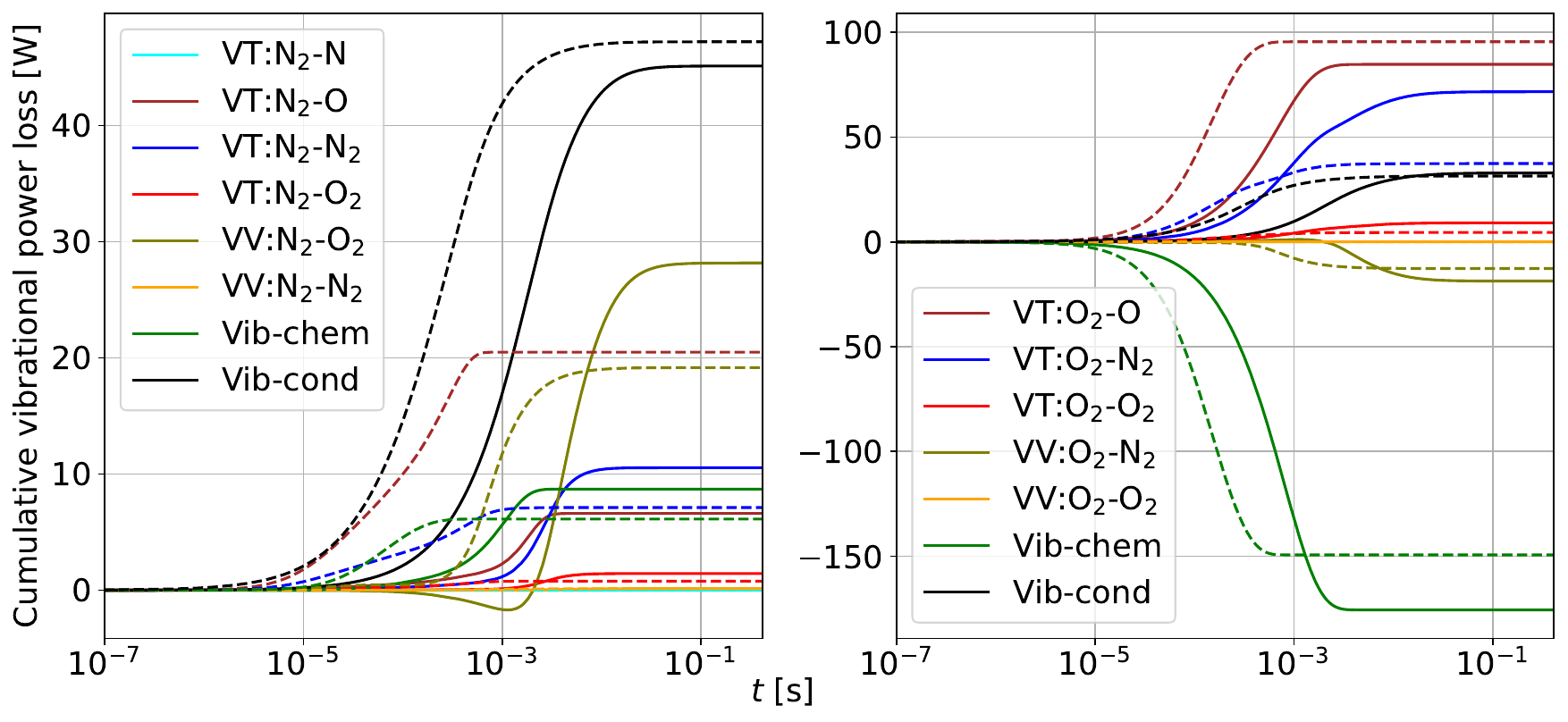}
\caption{The \secondround{cumulative} vibrational energy loss of N$_2$ (left) and O$_2$ (right) by different channels \fourround{over different time intervals} in the quenching region with $T_{CSTR}$= 3500 K for different cooling rates, \fourround{based on Eq.\ref{eq:vib reaction production}}. Solid: $c$ = 10; dash: $c$ = 100. The vib-chem term represents the total vibrational power loss contributed to all relevant chemical reactions. The vib-cond term denotes the vibrational power loss due to conduction.}
\label{vib_energy_with_time}
\end{figure}

The time-dependent evolution of vibrational energy losses \myworries{through different channels} with $T_{CSTR}$ = 3500~K at various cooling rates is illustrated in Figure~\ref{vib_energy_with_time}. Under the condition of a low cooling rate ($c = 10$), \myworries{most of the vibrational energy of N$_2$ is lost through conduction, accounting for 44.8\%}. Since V-V relaxation is less sensitive to gas temperature compared to V-T relaxation, it remains efficient for a longer duration. As a result, \myworries{the V-V N$_2$-O$_2$ process contributes nearly 28\% to the vibrational energy loss of N$_2$. Furthermore, 8.8\% of the vibrational energy is converted into chemical energy, which is slightly lower than the energy lost via the V-T N$_2$-O$_2$ process}. Despite the similar relaxation rate coefficients between the V-T N$_2(v)$-O$_2$ and N$_2$-N$_2$ processes (Figure \ref{fig: VVandVTrate}), the limited amount of O$_2$ restricts its contribution. \myworries{Likewise, although the V-T N$_2$-N  process exhibits the highest relaxation rate coefficient at the same temperature, its contribution is similarly restricted due to the limited availability of N atoms.} At a higher cooling rate ($c = 100$), \myworries{a larger fraction of the vibrational energy of N$_2$ is dissipated through heat conduction.} The onset of deactivation processes occurs approximately ten times faster, \myworries{leading to the V-T N$_2$-O process playing a more significant role, accounting for nearly 20\% of vibrational energy loss. Consequently, the contributions of the V-V N$_2$-N$_2$ process and the V-T N$_2$-N$_2$ process to vibrational energy loss decrease.}

\begin{figure}[t]
\centering
\includegraphics[width=1\linewidth]{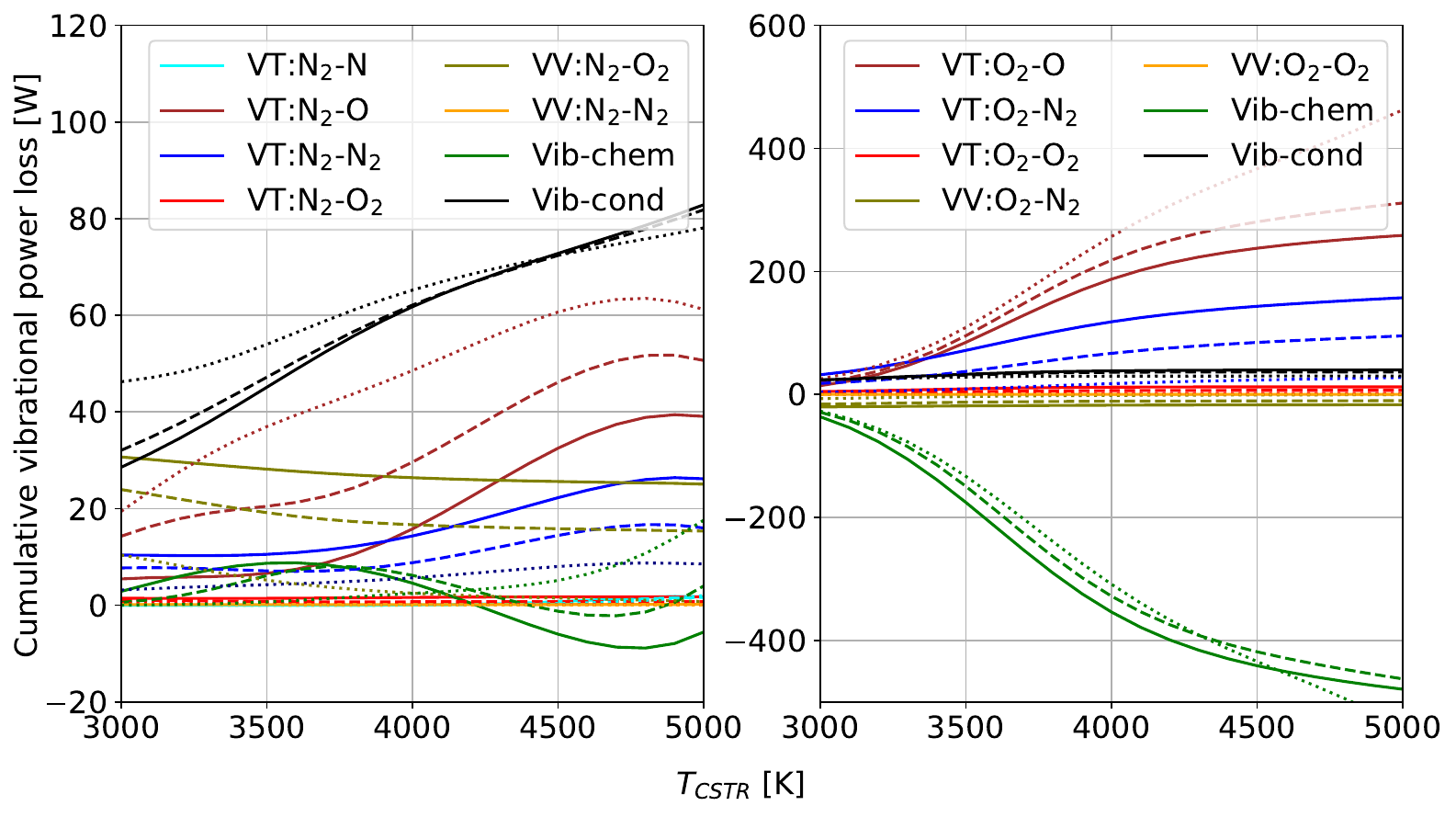}
\caption{ The \fourround{cumulative} vibrational energy loss of N$_2$ (left) and O$_2$ (right) by different channels \fourround{in the whole quenching process} as a function of $T_{CSTR}$ at different cooling rates, \fourround{based on Eq.\ref{eq:vib reaction production}}. Solid line: $c$ = 10; dash line: $c$ = 100; dot line: $c$ = \myworries{1000}. The vib-chem term represents the total vibrational power loss contributed to all relevant chemical reactions. The vib-cond term denotes the vibrational power loss due to conduction.}
\label{vib_N2_O2_as_T}
\end{figure}

\secondround{Chemical energy released by relevant chemical reactions plays a vital role in enhancing the vibrational energy of O$_2$ in the quenching region.} The V-T O$_2$-O process consistently dominates the vibrational deactivation of O$_2$ \secondround{across} different cooling rates, \secondround{especially} at higher cooling rates, whereas the V-T O$_2$($v$)-N$_2$ and O$_2$-O$_2$  processes exhibit the opposite trend  (Figure \ref{vib_energy_with_time}). Notably, a portion of the vibrational energy of N$_2$ can be converted into that of O$_2$ through the V-V process. and \secondround{the converted power} decreases with \secondround{higher} cooling rate. Because of \secondround{ anharmonicity}, V-V interactions of N$_2$-N$_2$ and O$_2$-O$_2$ result in some vibrational energy being lost, but the effects are limited (Figure~\ref{vib_energy_with_time}).

In addition to the cooling rate, $T_{CSTR}$ also influences the \secondround{contributions} of different relaxation processes (Figure \ref{vib_N2_O2_as_T}). With an increase \secondround{in} $T_{CSTR}$ in the plasma region, more recombination processes can occur in the afterglow, leading to more chemical energy \secondround{being converted} to vibrational energy. The increased number of O atoms in the plasma region at higher $T_{CSTR}$ also promotes the V-T N$_2$-O and O$_2$-O processes to become \secondround{one of} the dominant channels for the vibrational energy loss of N$_2$ and O$_2$, respectively. \myworries{Thermal conduction accounts for most of the vibrational energy loss of N$_2$ under most conditions but plays a limited role in the vibrational energy loss of O$_2$.} Although the V-T rate coefficients for N$_2$($v$)-O$_2$ and O$_2$-O$_2$ increase significantly with rising gas temperature, the limited availability of reactants constrains their \secondround{contributions} to vibrational energy relaxation, particularly at high temperatures. Since the V-V N$_2$-O$_2$ process primarily affects the vibrational energy exchange between N$_2$ and O$_2$ at low gas temperatures (Figure \ref{vib_energy_with_time}), their contribution is not sensitive to $T_{CSTR}$ and even slightly decreases with increasing $T_{CSTR}$. For the remaining deactivation processes, $T_{CSTR}$ has a limited impact on their contribution to vibrational energy loss. As the cooling rate increases, more vibrational energy of N$_2$ and O$_2$ is consumed by the O atoms (\(i.e.,\) by V-T N$_2$-O and O$_2$-O processes). However, the cooling rate plays the opposite role in the contributions of $Q_{VT}^{N_2-N_2}$, $Q_{VT}^{O_2-N_2}$, and $Q_{VV}^{N_2-O_2}$.

\subsection{Temporal variation of temperatures at different cooling rates}

\begin{figure}[ht]
\centering
\includegraphics[width=1\linewidth]{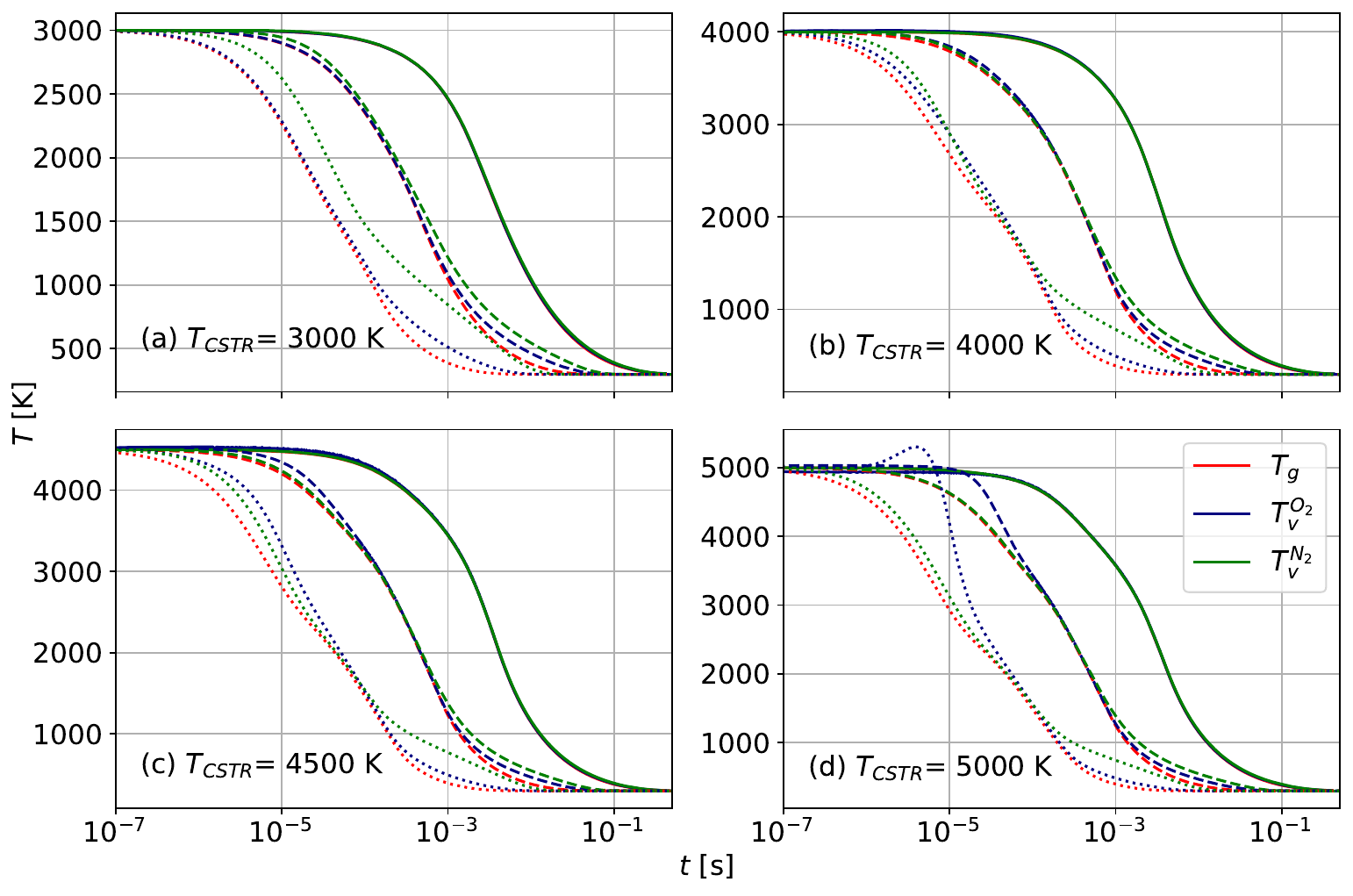}
\caption{Temporal variation of the temperatures in the quenching process at different cooling rates and $T_{CSTR}$. Solid line: $c$ = 10; dash line: $c$ = 100; dot line: $c$ = \myworries{1000}.}
\label{temperature_differnet_cooling_rate}
\end{figure}

Since temperature affects both the chemical and the vibrational relaxation \secondround{processes}, modelling \secondround{the} gas temperature is of major importance \cite{pintassilgo2016power}. Figure \ref{temperature_differnet_cooling_rate} reveals the temporal evolution of gas and vibrational temperatures. \myworries{\secondround{At a low cooling rate ($c$ = 10), all temperatures remain nearly in equilibrium throughout the entire quenching region.} It is because V-T relaxation is sufficiently efficient to match the wall heat losses in the gas heat balance, leading to vibrational temperatures close to the gas temperature.} Hence, it can be effectively replaced by \secondround{a} thermal quenching model, with much \secondround{lower} computational cost. As the cooling rate increases to $c$ = 100 or even \myworries{1000}, intensified cooling leads to a more rapid decline in gas temperature than \secondround{in} vibrational temperatures.

\secondround{At the beginning of the quenching process with $T_{CSTR}$ = 3000~K, the difference between $T_v^{O_2}$ and $T_g$ is limited, but it increases as $T_{CSTR}$ rises.} The reason is that \thirdround{less O$_2$} is dissociated in the plasma region at low $T_{CSTR}$, which confines the supply of the vibrational energy of O$_2$ through the recombination reactions. \myworries{Under the condition of $T_{CSTR}$ = 5000 K, nearly all O$_2$ dissociates into O atoms in the plasma region (Figure~\ref{Mole fraction and energy cost in hot zone}), resulting in a high abundance of O atoms at the onset of the afterglow. Due to the high cooling rate, the rapid temperature drop triggers the recombination of O atoms, which releases a significant amount of chemical energy that is subsequently converted into the vibrational energy of O$_2$. \fourround{Due to the limited availability of O$_2$ at elevated temperatures, even a minor portion of the recombination energy can significantly enhance the vibrational temperature of O$_2$ during the initial quenching phase, when} the vibrational energy generated from chemical energy exceeds that consumed by the V-T relaxation processes. This leads to an increase in the vibrational temperature of O$_2$ at the beginning of the afterglow under the conditions of high cooling rate and $T_{CSTR}$.}

\myworries{In contrast to O$_2$, the non-thermal behaviour of N$_2$ is more likely to occur under conditions of low $T_{CSTR}$ and high cooling rates. The vibrational temperature of N$_2$ drops rapidly at high $T_{CSTR}$ due to the deactivation effect by O atoms \cite{silva2024unraveling} through the V-T N$_2$-O process. As O atoms \secondround{disappear}, the decrease in the vibrational temperature of N$_2$ slows \secondround{during} the latter half of the quenching process. It is noteworthy that, although a portion of the vibrational energy of N$_2$ is transferred to O$_2$ via the V-V O$_2$-N$_2$ process in the later stages of quenching (Figure \ref{vib_energy_with_time}), conduction still limits the persistence of non-thermal behaviour in O$_2$. As a result, the vibrational temperature of O$_2$ is closer to the gas temperature compared to that of N$_2$.}

\subsection{Energy cost in the quenching process}

\begin{figure}[ht]
\centering
\includegraphics[width=1\linewidth]{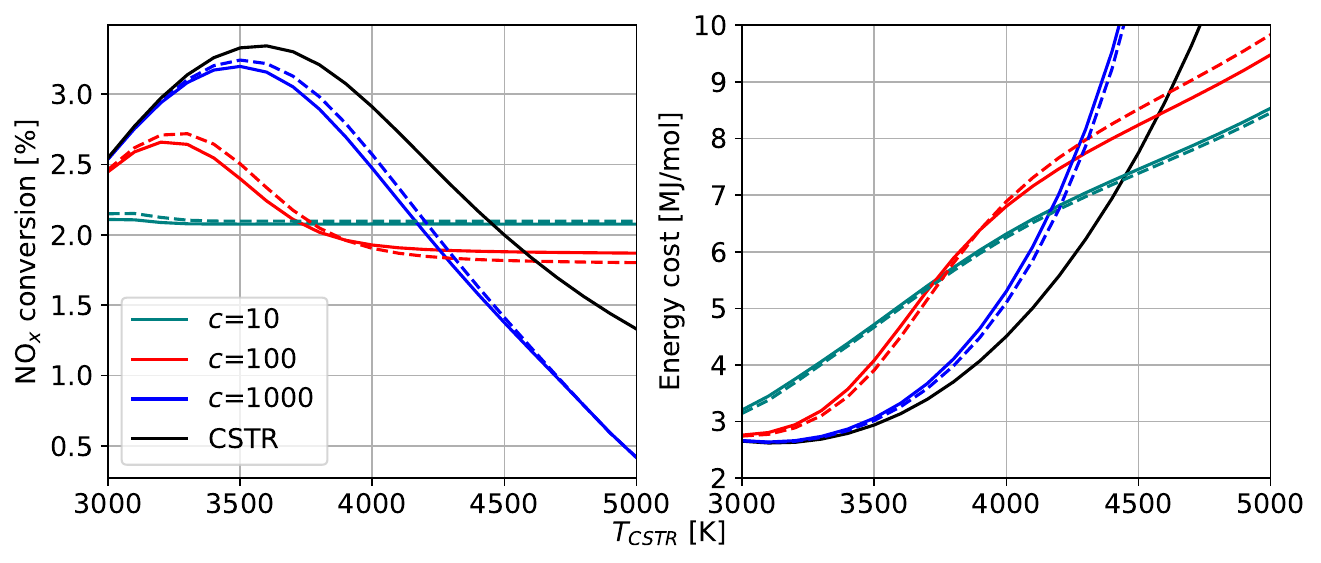}
\caption{NO$_x$ conversion and energy cost as a function of $T_{CSTR}$ at different cooling rates. Solid line: non-thermal model; dash line: thermal model (\(i.e.,\) with $T_g$ = $T_v^{N_2}$ = $T_v^{O_2}$ and the heat balance equation with full thermodynamic properties).}
\label{energy_cost}
\end{figure}

\begin{figure}[t]
\centering
\includegraphics[width=0.7\linewidth]{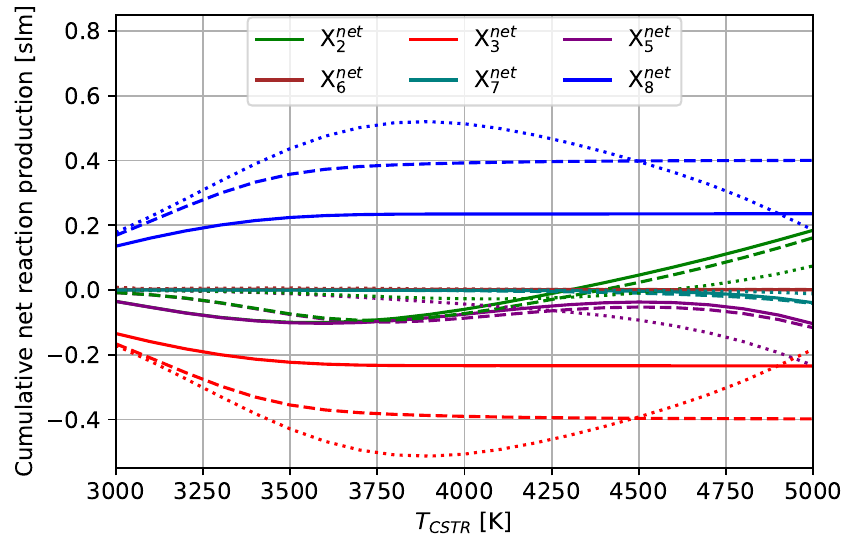}
\caption{\fourround{The cumulative net reaction production} related to NO and NO$_2$ formation in the whole quenching process as a function of $T_{CSTR}$ at different cooling rates, \fourround{based on Eq.\ref{Eq.reaction production}}. Solid line: $c$ = 10; dash line: $c$ = 100; dot line: $c$ = 1000. \fourround{X$^{net}$ represents that both forward and backward reactions are taken into account.} }
\label{Net reaction related to NO}
\end{figure}

In order to gain insight into the performance of the quenching process, \secondround{the} NO$_x$ conversion and \secondround{the} energy cost as a function of $T_{CSTR}$ at different cooling rates \secondround{are investigated} (Figure~\ref{energy_cost}). Unlike plasma-activated CO$_2$ conversion, which requires a higher cooling rate in the quenching process \cite{kim2020carbon}, NO$_x$ formation exhibits a more intricate relationship with the cooling rate. If $T_{CSTR}$ ranges between 3000~K and 4200~K, a fast quenching process ($c$ = \myworries{1000}) can increase NO$_x$ conversion and thereby decrease the energy cost. This improvement occurs because a high cooling rate can effectively reduce the detrimental effects of the reverse Zeldovich mechanism on  NO production in the quenching region \cite{abdelaziz2023toward} (Figure \ref{Net reaction related to NO}). As a result, \secondround{the highest conversion in the quenching region is 3.2\% at 3500 K, while the lowest energy cost is 2.6 MJ/mol N$^{-1}$ at 3110 K.} Furthermore, the non-thermal effect emerges during quenching at higher cooling rates (Figure \ref{energy_cost}), amplifying with increasing $T_{CSTR}$ due to higher vibrational energy transfer by recombination heating (Figure \ref{energy utilisation in hot region}). \secondround{A relatively strong vibrational enhancement of NO$_x$ production is observed at intermediate cooling rates, compared to low and high cooling rates. This is because a low cooling rate allows thermal equilibrium to be reached during quenching (Figure \ref{temperature_differnet_cooling_rate}), while a high cooling rate causes vibrational temperatures to drop too quickly due to conduction, limiting vibrational energy for NO production within the timescale of chemical reactions.}  Despite this enhancement, the energy cost under the non-thermal state still fails to break the thermal limit of \secondround{the} CSTR under absolute quenching conditions across different cooling rates. With the optimal $T_{CSTR}$ (\textit{i.e.,} 3110 K), fast quenching ($c$ = 1000) can preserve nearly all the NO$_x$ produced in the plasma region. However, due to the constrained recombination energy (Figure \ref{energy utilisation in hot region}), NO$_x$ production enhanced by the non-thermal effect is very limited at the same $T_{CSTR}$. Table~\ref{tab:energy_cost} reports the experimental energy cost \secondround{of} nitrogen fixation \secondround{in} (near) atmospheric warm air plasma \secondround{under} various conditions and plasma sources. The lowest energy cost found in all these experiments ($\sim$ 2.6 MJ/mol N$^{-1}$) is very similar to that predicted by the model presented under optimal conditions in this paper (Figure~\ref{energy_cost}).

\begin{table}
    \centering
    \small
    \begin{tabular}{C{5cm}C{1.4cm}C{1.6cm}C{4.5cm}C{0.5cm}}
    \hline
    Plasma type  & Pressure & NO$_x$ [\%] & Energy cost [MJ/mol N$^{-1}$] & Ref.\\
    \hline
    MW & 1 atm  & 3.0 & 2.6  &  \cite{kelly2021nitrogen}\\
    MW & $\approx$1 atm  & 0.45 & 3.76  &  \cite{kim2010formation}\\
    GA plasmatron & 1 atm &  1.4 & 3.8  & \cite{vervloessem2020plasma} \\
    Rotating GA & 1 atm & 2.5 & 3.0 & \cite{jardali2021no} \\
    Rotating GA & 1 atm & 0.8 &  2.7 & \cite{majeed2024effect} \\
    Rotating GA   & 1 atm & 3.9 &  3.5 & \cite{tsonev2023nitrogen} \\
    Magnetic field glow discharge   & 1 atm & 1.8 &  2.65 & \cite{li2022atmospheric} \\
    Magnetic field rotating GA   & 1 atm & 1.3 &  4.2 & \cite{chen2021highly} \\
    \hline
    \end{tabular}
    \caption{Experimental results of the nitrogen fixation yield and energy cost for different warm air (or N$_2$:O$_2$ = 4:1) plasma under (near) atmospheric pressure. The values quoted are obtained under various conditions.}
    \label{tab:energy_cost}
\end{table}

As the $T_{CSTR}$ exceeds \myworries{3600~K}, the NO loss through reverse Zeldovich reactions decreases \secondround{with} the low cooling rate ($c$= 10), while it continues to increase \secondround{with} higher cooling rates. Since the reaction rate coefficients of X$_2^f$ are similar to that of X$_5^b$ at high temperatures but much lower than that of X$_5^b$ at low temperatures, a low cooling rate can extend the high-temperature region and thus enhance the reaction time of N atoms with O$_2$. \secondround{Thus, with a low cooling rate, N atoms produced in the plasma region are able to combine with more O$_2$, which minimizes the destruction of NO within the quenching region.} Consequently, the energy cost \secondround{with} the low cooling rate ($c$ = 10) becomes lower than that \secondround{with} $c$ = 100 (at \myworries{3770~K}) and $c$ = \myworries{1000} (at \myworries{4200~K}), respectively. \secondround{When $T_{CSTR}$ exceeds 4450 K}, the higher NO production by the reactions X$_2^{net}$ and X$_7^{net}$, in comparison to the NO loss via the reaction X$_5^{net}$, boosts NO production at $c$ = 10 in the quenching process, \secondround{thereby reducing the energy cost after quenching}. However, the higher cooling rate results in lower net NO production via the reactions X$_2^{net}$, coupled with higher NO loss through the reactions X$_5^{net}$ and X$_7^{net}$. In summary, the optimal cooling rate for NO$_x$ formation depends strongly on $T_{CSTR}$. While fast quenching can enhance NO$_x$ production \secondround{for} moderate temperatures (3000\(-\) 4200 K), it becomes less beneficial \secondround{for} higher temperatures due to the reversed Zeldovich reactions. Consequently, careful optimisation of both $T_{CSTR}$ and cooling rate is essential to achieve efficient NO$_x$ production and thus reduce energy cost. 

\secondround{The model indicates that non-thermal effects have a limited influence on the afterglow of air plasma at atmospheric pressure. However, it is expected to be more applicable at lower pressures. In MW plasma, where the reduced electric field typically ranges from 10 to 100 Td \cite{viegas2020insight}, most electron energy is transferred to the vibrational excitation of N$_2$ \cite{li2023magnetic}. As pressure decreases, the impact of V-T relaxation processes weakens, leading to more pronounced non-thermal effects. Considering that strong non-thermal effects have been observed in the experiment at low pressures \cite{bahnamiri2021nitrogen}, the non-thermal effects may also persist in the quenching region, potentially enhancing NO production. Furthermore, diffusion effects become more significant at lower pressures, resulting in less steep temperature gradients \cite{wolf2019characterization}, making the 1D axial model a more suitable approximation for such conditions.}

\begin{figure}[t]
\centering
\includegraphics[width=1\linewidth]{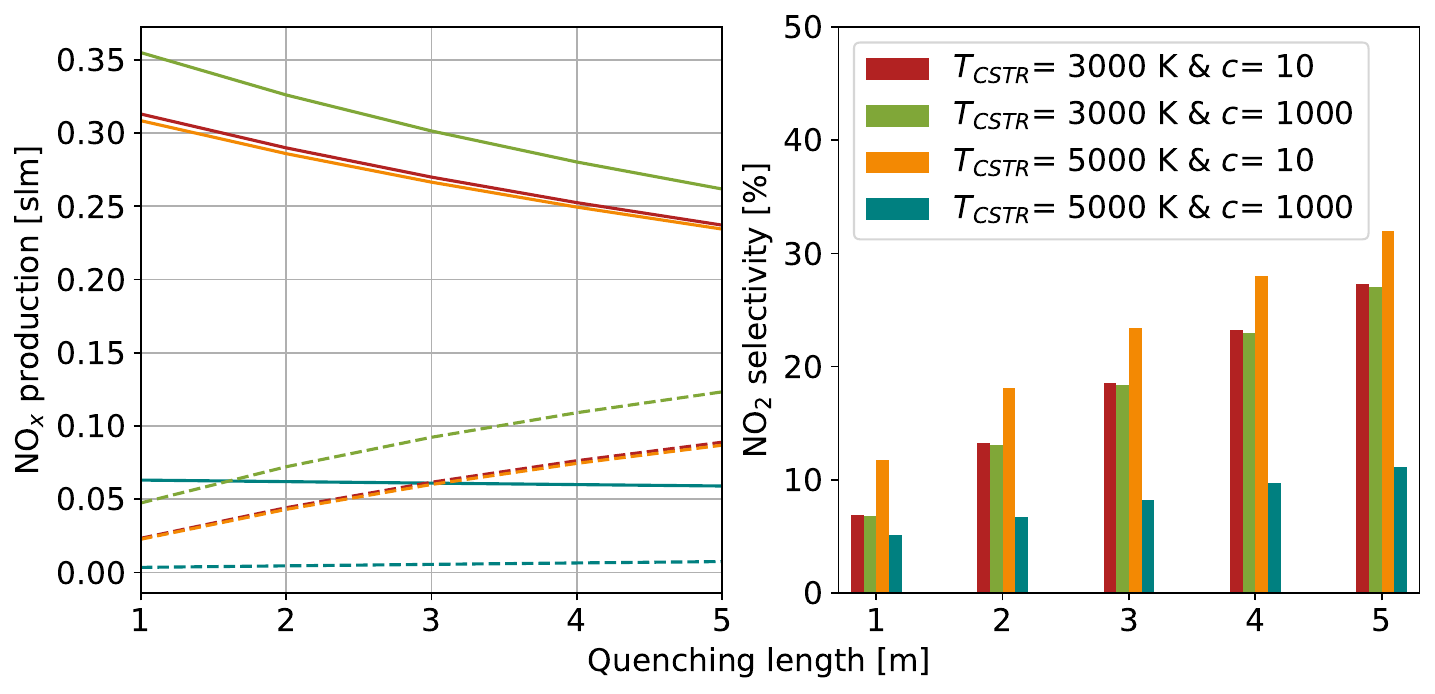}
\caption{NO$_2$ selectivity and NO$_x$ production as a function of quenching length under different conditions. Solid line: NO production; dash line: NO$_2$ production; column data: NO$_2$ selectivity.}
\label{fig: NO2_selectivity}
\end{figure}

Besides NO, NO$_2$ is another product of interest, primarily affected \secondround{by} reactions X$_3^{net}$, X$_6^{net}$, and X$_8^{net}$ in this study. As previously discussed, the high temperature within the CSTR significantly limits NO$_2$ production (Figure \ref{Mole fraction and energy cost in hot zone}). At the beginning of the quenching process, where most recombination processes take place, nearly all the NO$_2$ generated by reaction X$_8^f$ is consumed by reaction X$_3^b$ (Figure \ref{Net reaction related to NO}), providing an effective pathway for oxygen recombination. Nearly all NO$_2$ forms in the later part of the quenching process, primarily facilitated by reaction X$_6^b$, which is relatively slow and operates optimally at lower gas temperatures. The production and selectivity of NO$_2$ as a function of quenching length under various conditions are illustrated in Figure \ref{fig: NO2_selectivity}. Notably, NO$_2$ selectivity \secondround{exhibits a positive correlation} with residence time. The influence of the cooling rate and $T_{CSTR}$ on NO$_2$ production is less pronounced compared \secondround{to that of} residence time. Furthermore, under conditions of high $T_{CSTR}$ and high cooling rate ($T_{CSTR}$ = 5000~K and $c$ = \myworries{1000}), the limited NO production constrains reaction X$_6^{b}$, leading to a lower NO$_2$ selectivity. Given that reaction X$_6^b$ is exothermic, its progression does not require additional thermal excitation. Therefore, with an infinitely long residence time, all NO will eventually convert to NO$_2$ if sufficient O$_2$ is available, reaching chemical equilibrium. This suggests that the key to maximizing NO$_2$ production lies in extending the residence time to ensure sufficient reaction time for NO$_2$ formation processes, particularly at lower temperatures in the quenching region where reaction X$_6^b$ is most effective.

\section{Conclusions}

\secondround{The primary aim of this study is to investigate} the energy transfer mechanisms and NO$_x$ formation \secondround{that occur} during the quenching process of dry air MW plasma at atmospheric pressure. A 1D multitemperature chemical kinetics model that \secondround{incorporates a} common translational-rotational temperature and \secondround{separate} vibrational temperatures for N$_2$ and O$_2$, connected \secondround{through} various vibrational excitation and relaxation reactions, \secondround{is introduced}. \secondround{Given the inherent limitations of the 1D model, it is incapable of simulating temperature gradients in the radial direction. Consequently, the primary objective of this study is not to replicate exact experimental data but to \fourround{serve as a crucial first step toward optimizing plasma reactors for efficient NO$_x$ production.}}

Firstly, the properties of the thermal plasma region \secondround{is analysed using the CSTR model, demonstrating} that chemical equilibrium is reached at typical plasma lengths. \secondround{It is observed} that the lowest energy cost \secondround{in} the plasma region is 2.6~MJ/mol N$^{-1}$ at \secondround{a} gas temperature in the plasma region of $T_{CSTR}$ = 3110 K. However, more than 81\% of the total energy is dissipated for gas heating at the same time. 

The downstream model shows that the optimal cooling rate in the quenching region depends on $T_{CSTR}$. At relatively low $T_{CSTR}$ values (3000–4200 K), a higher cooling rate enhances O$_2$ recombination, which reduces NO loss through the reverse Zeldovich mechanism. Consequently, the optimal conditions for NO formation \secondround{are found} at the optimal $T_{CSTR}$ (3110 K) and with rapid quenching (characterized \secondround{by} $c$ = \myworries{1000}). However, due to the limited recombination energy at the optimal $T_{CSTR}$, the non-thermal effects during quenching cannot further facilitate NO formation or \secondround{overcome} the thermal limit. At higher $T_{CSTR}$ values (above 4200 K), a lower cooling rate allows the mixture to remain longer in the optimal temperature range for NO formation, which \secondround{can potentially} reduce the energy cost. \secondround{Therefore, at a low cooling rate, more N atoms can combine with O$_2$ rather than reacting with NO, which occurs at a higher cooling rate.} \myworries{Compared to low and high cooling rates, a greater vibrational enhancement of NO$_x$ production occurs under intermediate cooling rate conditions.} \secondround{Although non-thermal effects have a limited influence on the afterglow of air plasma at atmospheric pressure, we believe the application of this model to lower-pressure conditions can bring invaluable insight, where non-thermal effects are more pronounced.}

\myworries{Thirdly, the impact of various vibrational energy loss processes of N$_2$ and O$_2$ in the quenching region is analyzed. The predominant factors limiting vibrational non-equilibrium for N$_2$ and O$_2$ are \secondround{vibrational} heat conduction and the V-T O$_2$-O process, respectively.} As $T_{CSTR}$ increases, aided by higher V-T relaxation rates and more \secondround{dissociated oxygen} in the plasma region, V-T O$_2$-O and N$_2$-O processes consume more vibrational energy of O$_2$ and N$_2$. The V-V N$_2$-O$_2$ process undertakes a more significant role in the vibrational energy exchange of N$_2$ with low $T_{CSTR}$. \secondround{The model also predicts how different temperatures evolve over time} under varying cooling rates during the quenching process.

Finally, the NO$_2$ formation in the plasma and quenching regions \secondround{is discussed,} separately. Due to the higher rate coefficients of the reaction NO$_2$+O, all the NO$_2$ formed by the reaction of NO+O+M is consumed by O atoms in the plasma region and the early part of the quenching region, where $T_g$ remains high. \secondround{The model} results highlight the reaction of 2NO+O$_2$ as the primary driver of NO$_2$ formation. This reaction requires sufficient residence time and low gas temperature to proceed effectively.

\section{Acknowledgement}
This work was financially supported by the China Scholarship Council (Grant No. CSC202106240037). VG was partially supported by FCT – Fundação para a Ci\^encia e Tecnologia under the projects UIDB/50010/2020 \\(https://doi.org/10.54499/UIDB/50010/2020), \\UIDP/50010/2020 (https://doi.org/10.54499/UIDP/50010/2020), \\LA/P/0061/2020 (https://doi.org/10.54499/LA/P/0061/2020), and \\PTDC/FIS-PLA/1616/2021 (https://doi.org/10.54499/PTDC/FIS-PLA/1616/2021). \thirdround{The author sincerely thanks Lex Kuijpers for their valuable assistance in polishing the English language of this article.}
\newpage
\section{Supplementary material}

\renewcommand{\thefigure}{S1}
\begin{figure}[h]
\centering
\includegraphics[width=0.6\linewidth]{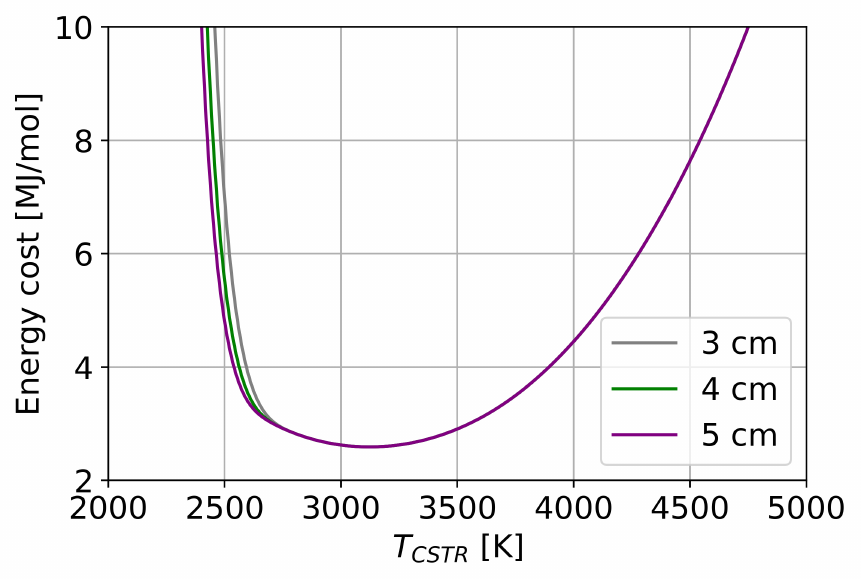}
\caption{Energy cost in the plasma length as a function of $T_{CSTR}$ at different plasma lengths under atmospheric pressure.}
\label{fig:S1}
\end{figure}

\renewcommand{\thefigure}{S2}
\begin{figure}[h]
\centering
\includegraphics[width=1\linewidth]{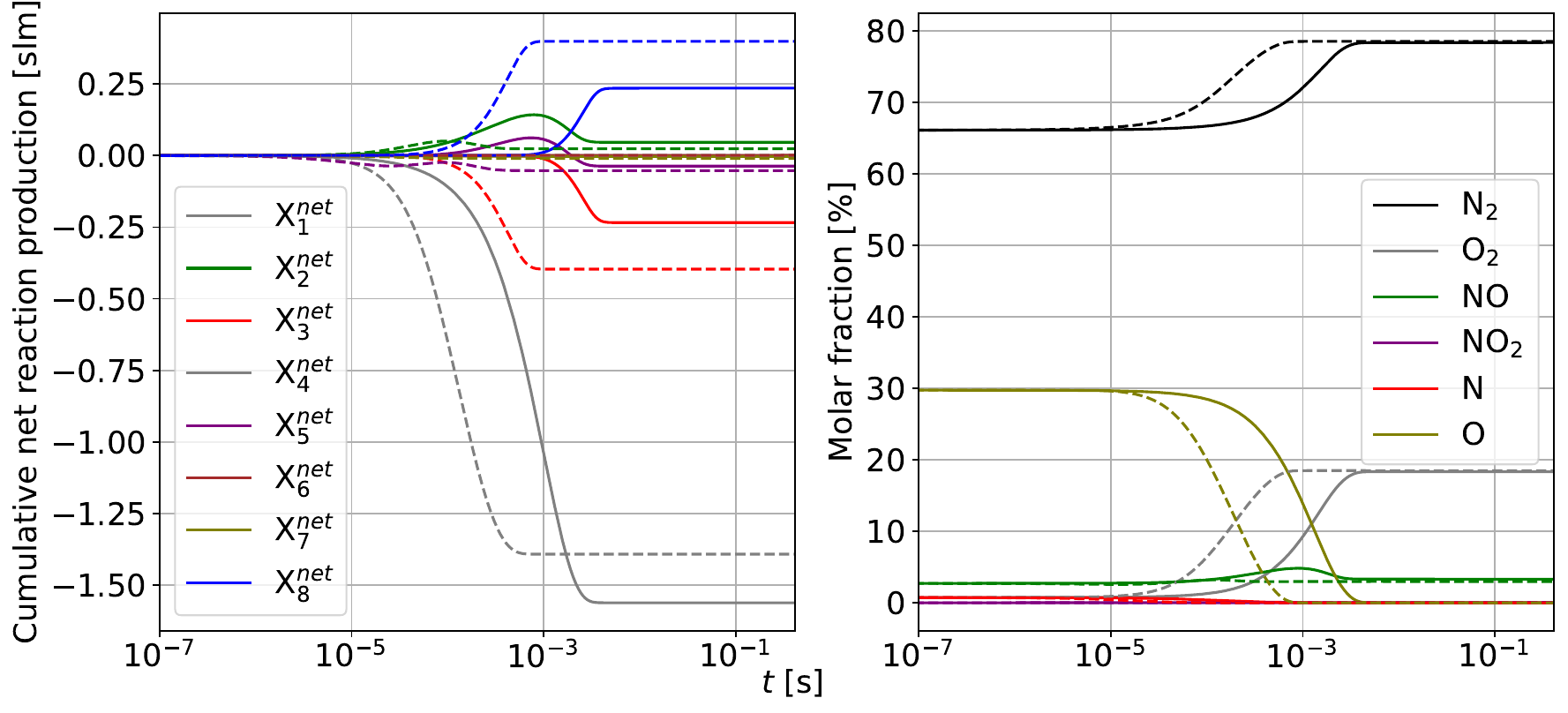}
\caption{The net reaction production and mole fractions as a function of the time in the quenching region with $T_{CSTR}$ = 4500~K at different cooling rates. Solid: $c$ = 10; dash: $c$ = 100. X$^{net}$ means considering both forward and backward reaction effects.}
\label{fig:S2}
\end{figure}

\newpage
\renewcommand{\thetable}{S1}

\begin{table}[h]
\caption{O$_2$ vibrational levels and the corresponding energies at the respective ground rotational levels \cite{esposito2008o2}.}
\centering
\begin{tabular}{cccccccccc}
 \hline
 \(v\) & E$_v$ (eV) & \(v\) & E$_v$ (eV) & \(v\) & E$_v$ (eV) & \(v\) & E$_v$ (eV) & \(v\) & E$_v$ (eV) \\
 \hline
0 & 0.09745 & 10 & 1.03779 & 20 & 2.99673 & 30 & 4.52001 & 40 & 5.09830 \\
1 & 0.29065 & 11 & 1.21788 & 21 & 3.13813 & 31 & 4.60590 & 41 & 5.13293 \\
2 & 0.48125 & 12 & 1.39517 & 22 & 3.27583 & 32 & 4.68674 & 42 & 5.16100 \\
3 & 0.66941 & 13 & 1.56956 & 23 & 3.40973 & 33 & 4.76239 & 43 & 5.18243 \\
4 & 0.85490 & 14 & 1.74104 & 24 & 3.53983 & 34 & 4.83271 & 44 & 5.19735 \\
5 & 1.03779 & 15 & 1.90934 & 25 & 3.66583 & 35 & 4.89753 & 45 & 5.20638 \\
6 & 1.21788 & 16 & 2.07464 & 26 & 3.78783 & 36 & 4.95668 & 46 & 5.21082 \\
7 & 1.39517 & 17 & 2.23674 & 27 & 3.90563 & 37 & 5.00998 & & \\
8 & 1.56956 & 18 & 2.39554 & 28 & 4.01922 & 38 & 5.05725 & & \\
9 & 1.74104 & 19 & 2.55104 & 29 & 4.12852 & 39 & 5.09830 & & \\
 \hline
\end{tabular}
\end{table}

\renewcommand{\thetable}{S2}
\begin{table}[h]
\caption{N$_2$ vibrational levels and the corresponding energies at the respective ground rotational levels \cite{esposito2017reactive}.}
\centering
\begin{tabular}{cccccccccc}
\hline
\(v\) & \(E_v\) (eV) & \(v\) & \(E_v\) (eV) & \(v\) & \(E_v\) (eV) & \(v\) & \(E_v\) (eV) & \(v\) & \(E_v\) (eV) \\
\hline
0 & 0.151726 & 13 & 3.605168 & 26 & 6.367745 & 39 & 8.410388 & 52 & 9.646625 \\
1 & 0.441947 & 14 & 3.842309 & 27 & 6.551098 & 40 & 8.535777 & 53 & 9.703346 \\
2 & 0.728036 & 15 & 4.075379 & 28 & 6.730192 & 41 & 8.656395 & 54 & 9.753875 \\
3 & 1.010007 & 16 & 4.304370 & 29 & 6.905002 & 42 & 8.772182 & 55 & 9.797986 \\
4 & 1.287873 & 17 & 4.529277 & 30 & 7.075500 & 43 & 8.883070 & 56 & 9.835412 \\
5 & 1.561646 & 18 & 4.750090 & 31 & 7.241657 & 44 & 8.988990 & 57 & 9.865829 \\
6 & 1.831331 & 19 & 4.966799 & 32 & 7.403441 & 45 & 9.089864 & 58 & 9.888824 \\
7 & 2.096939 & 20 & 5.179395 & 33 & 7.560820 & 46 & 9.185609 & 59 & 9.903833 \\
8 & 2.358474 & 21 & 5.387863 & 34 & 7.713757 & 47 & 9.276135 & 60 & 9.913001 \\
9 & 2.615942 & 22 & 5.592189 & 35 & 7.862214 & 48 & 9.361342 & & \\
10 & 2.869344 & 23 & 5.792359 & 36 & 8.006151 & 49 & 9.441121 & & \\
11 & 3.118683 & 24 & 5.988354 & 37 & 8.145524 & 50 & 9.515353 & & \\
12 & 3.363958 & 25 & 6.180156 & 38 & 8.280286 & 51 & 9.583904 & & \\
\hline
\end{tabular}
\end{table}
\newpage

\renewcommand{\thefigure}{S3}
\begin{figure}[h]
\centering
\includegraphics[width=1\linewidth]{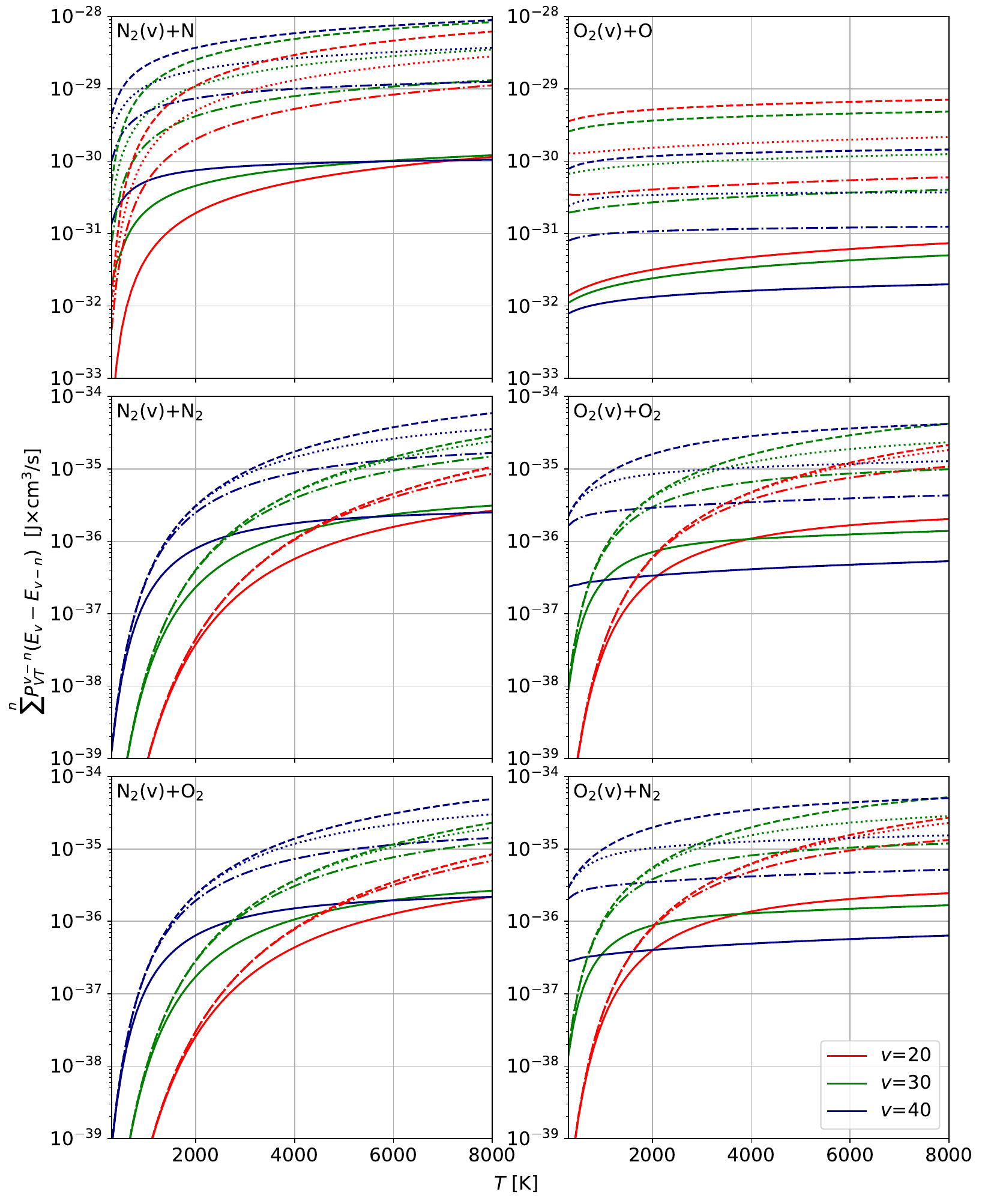}
\caption{Comparison of the energy transfer rates including V-T processes with different maximal numbers of quanta. Solid line: $n$ = 1 (\textit{i.e.,} only single quantum V-T relaxation is considered); dash-dotted line: $n$ = 5; dotted line: $n$ = 10; dash line: $n$ = 20.}
\label{fig:S3}
\end{figure}

\newpage


\bibliographystyle{iopart-num}
\bibliography{main_text}
\end{document}